\documentstyle[12pt]{article}
\begin {document}

\makeatletter
\@addtoreset{equation}{section}

\makeatother
\makeatletter

\baselineskip=24pt
\begin{center}{\Large \bf Local Systems of Vertex Operators,
 Vertex Superalgebras and Modules}\end{center}
\vspace{1.5cm}
\begin{center}{\bf Hai-sheng Li}\\
{\bf Department of Mathematics}\\
{\bf Rutgers University}\\
{\bf New Brunswick, NJ 08903}\end{center}
\vspace{1cm}
\begin{abstract}
We give an analogue for vertex operator algebras and superalgebras of
the notion of endomorphism ring of a vector space by
means of a notion of ``local system of vertex operators'' for
a (super) vector space. We first prove
that any local system of vertex operators on a (super)
vector space $M$ has a natural vertex (super)algebra structure
with $M$ as a module. Then we
prove that for a vertex (operator) superalgebra $V$, giving a
$V$-module $M$ is equivalent to giving
a vertex (operator) superalgebra homomorphism from $V$ to some local
system of vertex operators on $M$. As
applications, we prove that certain lowest weight modules for some well-known
infinite-dimensional Lie algebras or Lie superalgebras
have natural vertex operator superalgebra
structures. We prove the rationality of vertex operator superalgebras
associated to standard modules for an affine algebra.
We also give an analogue of the notion of the space of linear
homomorphisms from one module to another for a Lie algebra by
introducing a notion of ``generalized intertwining operators.'' We prove
that $G(M^{1},M^{2})$, the space of generalized intertwining
operators from one module $M^{1}$
to another module $M^{2}$ for a vertex operator superalgebra $V$, is a
generalized $V$-module.
Furthermore, we prove that for a fixed vertex operator superalgebra $V$ and
three $V$-modules $M^{i}$ $(i=1,2,3)$, giving an intertwining operator of
type $\left(\begin{array}{c}M^{3}\\M^{1},M^{2}\end{array}\right)$ is
equivalent to giving a $V$-homomorphism from $M^{1}$ to $G(M^{2},M^{3})$.
\end{abstract}

\newpage

\section{Introduction}
Vertex (operator) algebras ([B], [FLM]), which are
mathematical counterparts of chiral algebras in conformal field
theory (cf. [BPZ]), are analogues of both Lie algebras
and commutative associative algebras (see for example [B], [FHL], [FLM], [H]).
In classical Lie theory, both the notion of endomorphism algebra of a
vector space and the notion of space of linear maps between vector spaces
are fundmental and important. Thus we naturally
expect to have certain analogues for vertex operator algebras, which
may play important roles.

In this paper, we give an analogue for vertex operator superalgebras of the
notion of endomorphism algebra of a vector
space and an analogue of the notion of the space of linear
homomorphisms from one
module to another for a Lie algebra. As applications, we prove that
certain lowest weight modules  for certain
infinite-dimensional Lie algebras or Lie superalgebras such as the Virasoro
algebra, the Neveu-Schwarz algebra and affine superalgebras have
natural vertex operator superalgebra structures. We also prove the
rationality of certain  vertex operator superalgebras
associated to standard modules for an affine Lie superalgebra.
This gives an
alternate proof for some known results ([FZ], [Lian], [KW]).
Our results are more general than the corresponding results of [KW].

In vertex operator superalgebra theory, in addition to the notion of
module we have the notion of twisted module ([D2], [FFR], [FLM]).
All the corresponding results  of this paper in the twisted case have also been
obtained [L3].

Now we give an outline of this paper. For simplicity, let
us just consider the ordinary (``unsuper'') case; all the results
summarized here are carried out for superalgebras in this paper.
Starting with a vector space $M$, we consider
the vector space $({\rm End}M)[[z,z^{-1}]]$ consisting
of all formal series of operators on $M$, which can be viewed as a
complex analogue of ${\rm End}M$. Next, we need to find an appropriate
``algebra product.'' Suppose that a subspace $A$
of $({\rm End}M)[[z,z^{-1}]]$ equipped with a linear
map $Y(\cdot,z_{1})$ from $A$ to $({\rm End}A)[[z_{1},z_{1}^{-1}]]$ such that
$(A,Y)$ is a vertex algebra with $(M,Y_{M})$ as a natural
$A$-module, in the sense that $Y_{M}(a(z),z_{1})=a(z_{1})$ for any $a(z)\in A$.
Then
$Y_{M}(\cdot,z_{1})$ satisfies the Jacobi identity.
Let $a(z)$ and $b(z)$ be two elements of $A$.
Taking ${\rm Res}_{z_{1}}$ of the Jacobi identity for
$Y_{M}(\cdot,z_{1})$ and using the naturality, we obtain the following
``iterate formula:''
\begin{eqnarray}
& &Y_{M}(Y(a(z),z_{0})b(z),z_{2})\nonumber\\
&=&{\rm Res}_{z_{1}}\left(z_{0}^{-1}\delta\left(\frac{z_{1}-z_{2}}{z_{0}}
\right)a(z_{1})b(z_{2})-(-1)^{|a(z)||b(z)|}z_{0}^{-1}\delta\left(\frac{z_{2}
-z_{1}}{-z_{0}}\right)b(z_{2})a(z_{1})\right).
\end{eqnarray}
Since $c(z)=c(z_{1})|_{z_{1}=z}=Y_{M}(c(z),z_{1})|_{z_{1}=z}$ for any
$c(z)\in A$, we obtain
\begin{eqnarray}
Y(a(z),z_{0})b(z)=Y_{M}(Y(a(z),z_{0})b(z),z_{2})|_{z_{2}=z}.
\end{eqnarray}
Therefore
\begin{eqnarray}
& &Y(a(z),z_{0})b(z)\nonumber\\
&=&{\rm Res}_{z_{1}}\left(z_{0}^{-1}\delta\left(\frac{z_{1}-z}{z_{0}}
\right)a(z_{1})b(z)-(-1)^{|a(z)||b(z)|}z_{0}^{-1}\delta\left(\frac{z
-z_{1}}{-z_{0}}\right)b(z)a(z_{1})\right).
\end{eqnarray}
This is our desired ``product'' formula.
Noticing that the expression of the right hand side of (1.3) is not
well defined for arbitrary $a(z),b(z)\in ({\rm End}M)[[z,z^{-1}]]$, we
restrict our attention to the subspace $F(M)$
consisting of those formal series $a(z)$ satisfying the truncation
condition, i.e., $a(z)u\in M((z))$ for any $u\in M$; note that we must
have $A\subseteq F(M)$.
Then $ F(M)$ is a
subspace on which the right-hand side of (1.3) is well defined  and on
which $\displaystyle{D={d\over dz}}$ acts. Let
$I(z)={\rm id}_{M}$ be the identity operator of $M$. Then we obtain a
quadruple $(F(M), I(z),D,Y)$. But
$(F(M),D, I(z),Y)$ is not a vertex algebra.

To a certain extent,
vertex algebras look like commutative associative algebras with
identity because of the ``commutativity'' and the ``associativity.''
Just as there is no canonical largest commutative associative
algebra associated to a vector space,
there is no (canonical) universal or largest vertex
algebra associated to $M$, but there are maximal vertex
algebras inside the space $F(M)$, what we call ``local systems of
vertex operators.''

Formal series $a(z)$ and $b(z)$ of operators
on $M$ are
said to be {\it mutually local} if there is a positive integer $k$ such that
\begin{eqnarray}
(z_{1}-z_{2})^{k}a(z_{1})b(z_{2})=(z_{1}-z_{2})^{k}b(z_{2})a(z_{1}).
\end{eqnarray}
This is a variant form of the ``commutativity'' ([FLM], [FHL], [DL]) or
the ``locality'' (cf. [Go]). This variant seems to have been
first noticed and exploited by Dong and Lepowsky [DL]. An element
$a(z)$ of $F(M)$ is called a {\it (local) vertex operator} if $a(z)$
is local with itself.
Motivated by the notion of ``local system'' in [Go], we
define {\it a local subpace} of $F(M)$ to be a
subspace $A$ such that any two elements of $A$
are mutually local and we define {\it
a local system of vertex operators on $M$} to be a maximal local
subspace of $F(M)$. Then our first main result is that any local
system of vertex operators on $M$ is a vertex algebra
with $M$ as a module (Theorem 3.2.10).

To do this, we first prove that any local system $A$ of
vertex operators on $M$ is closed under the ``product'' (1.3)
(Proposition 3.2.7).
Then we prove that the quadruple $(A, I(z), D,Y)$ satisfies
the commutativity formula (1.4) (Proposition 3.2.9) and all the
axioms for vertex algebra except the Jacobi identity.
Then by proving
that these properties imply the Jacobi identity (Proposition 2.2.4),
we prove that any local system of vertex operators on $M$ is a
vertex algebra with $M$ as a module.

It has been well known ([FLM], [Go], [FHL], [DL]) that the ``commutativity''
implies the Jacobi identity for vertex operator algebra in the presence
of some elementary properties.
As one of their results, Dong and Lepowsky [DL] have proved that the
Jacobi identity  can be replaced by the
commutativity (1.4) for ``generalized vertex operator algebra''
and for ``generalized vertex algebra''
under a milder assumption. This refines and simplifies the
corresponding results of [FHL] and [FLM]. (See for example
[Guo] and [A] for other applications of formula (1.4).)

In the present paper, we refine DL's result further as follows:
Notice that the commutativity formula (1.4) is really a
commutativity formula for ``left multiplications.''
At the classical level, if $A$ is an algebra with identity such that the
left multiplications associated to any two elements of $A$ commute, then
one can easily prove that $A$ is a commutative associative algebra.
Motivated by this classical analogue,
instead of using the commutativity for products
of three vertex operators applied to ${\bf 1}$ as the intermediate
step as in [FHL] and [DL], we
first prove the skew-symmetry and then use this to prove the associativity.
This new route enables us to refine the corresponding result of [DL]
further for (generalized) vertex algebra (where the milder assumption
assumed in [DL] has been removed). The main difference
between vertex operator algebras and vertex algebras is that
two grading-restriction axioms are removed. For instance, there may be
no grading at all for an arbitrary vertex algebra.
Our proof is completely ``matrix coefficient''-free, so that it meets
our need for dealing with local systems without assuming the existence of
a certain grading.

Let $M$ be any vector space and let $S$ be any set of mutually
local vertex operators on $M$.
It follows from Zorn's lemma that  there exists a local
system $A$ containing $S$. Let $\langle S\rangle$ be the vertex
subalgebra of $A$ generated by $S$. Since the ``product'' (1.3) does not
depend on the choice of $A$, the vertex algebra $\langle S\rangle$ is
canonical and $M$ is a  $\langle S\rangle$-module (Corollary 3.2.11).
Let $V$ be a vertex (operator) algebra. Then we prove that
giving a $V$-module $M$ is equivalent to giving a vertex algebra
homomorphism from $V$ to some local system of vertex operators on $M$
(Proposition 3.2.13).

For applications, let us consider the following illustrative
example. Let $Vir$ be the Virasoro algebra and let $M$ be a
any restricted $Vir$-module, i.e., for any $u\in M$, $L(n)u=0$ for $n$
sufficiently large. Then $L(z)=\sum_{n\in {\bf Z}}L(n)z^{-n-2}$ can be
viewed as an element of $F(M)$.
Since the commutator formula implies the commutativity or
locality, the generating function $L(z)$ is a local
vertex operator on $M$. Then our results give a vertex algebra $V$
generated by $L(z)$ with $M$
as a $V$-module. One can prove that $V$ is a lowest weight
$Vir$-module with $I(z)$ as a lowest weight vector. (This module is
called the {\it vacuum module} by physicists.) More generally, as long
as a given ``algebra'' can be defined in
terms of the general ``cross bracket'' formula (2.2.6), this procedure
works. The main difference between Frenkel and Zhu's approach [FZ] and
ours is that they start with an ``algebra'' (a certain completion of the
universal algebra) while we start with a ``module.''

Let ${\bf g}$ be a finite-dimensional simple Lie algebra and let
$\tilde{{\bf g}}$ be the affine algebra.
let $\ell$ be a positive integer and let $L_{{\bf g}}(\ell,0)$ be the
standard $\tilde{{\bf g}}$-module of level $\ell$ with lowest
weight zero. As in the
previous paragraph, it is easy to see that $L_{{\bf g}}(\ell,0)$ is a
vertex operator algebra. (This has also been known in [FZ] and [DL].)
In [DL], Dong and Lepowsky have obtained an important formula
(Proposition 13.16 [DL]): $Y_{M}((v_{-1})^{k+1}{\bf 1},z)=Y_{M}(v,z)^{k+1}$,
where $M$ is a module for a vertex operator algebra $V$ and $v$ is an
element of $V$ such that $[Y_{M}(v,z_{1}),Y_{M}(v,z_{2})]=0$, so that
$Y_{M}(a,z)^{k}$ is well defined for any $k\in {\bf N}$.
Using this formula they
have proved that every irreducible $L_{{\bf g}}(\ell,0)$-module
is a standard $\tilde{{\bf g}}$-module (of level $\ell$) (Proposition
13.17 [DL]). For the other direction, they
have noticed that every standard $\tilde{{\bf
g}}$-module of level $\ell$ which can be obtained from the tensor
product of $\ell$
basic standard $\tilde{{\bf g}}$-modules (of type $A, D$ or $E$) is an
irreducible module for the vertex operator algebra.
(See [Xie] for further developments in this direction.)

In this paper, combining the complete reducibility of certain integrable
$\tilde{{\bf g}}$-modules with the complete reducibility of certain
${\bf g}$-modules, we prove that every lower
truncated ${\bf Z}$-graded ``weak'' $L_{{\bf g}}(\ell,0)$-module (relaxing one
of the two restrictions on the homogeneous subspaces) is a direct sum
of standard $\tilde{{\bf g}}$-modules of level $\ell$ (Proposition 5.2.6).
That is, $L_{{\bf g}}(\ell,0)$ is rational in the sense of [Z]. (This
extends DL's corresponding result.) Conversely, by employing
DL's formula and the analogue of endomorphism ring, we prove that
every standard $\tilde{{\bf g}}$-module of level $\ell$ is an $L_{{\bf
g}}(\ell,0)$-module (Proposition 5.2.4). These results have also been
obtained by Frenkel and Zhu [FZ] by using Zhu's $A(V)$-theory [Z].
Although our approach is closely related to FZ's approach, it has a
different flavor and it is also self-contained.

Let $M^{1}$ and $M^{2}$ be two modules for a
vertex operator algebra $V$. Then we introduce a notion of
what we call ``generalized intertwining operators'' from $M^{1}$ to
$M^{2}$ and prove
that ${\rm G}(M^{1},M^{2})$, the space of generalized
intertwining operators from $M^{1}$ to $M^{2}$, is a
generalized $V$-module. Furthermore, we prove that for any third
$V$-module $M$, giving an intertwining operator of type $\left(\begin{array}{c}
M^{2}\\M,M^{1}\end{array}\right)$ is equivalent to giving a
$V$-homomorphism from $M$ to ${\rm G}(M^{1},M^{2})$. Therefore,
we may think of ${\rm G}(M^{1},M^{2})$ as an analogue of the notion of
the space of
linear homomorphisms from one module to another for a Lie algebra.

This paper is organized as follows: In Section 2
we recall some basic definitions and prove that the Jacobi identity
for vertex superalgebra can be equivalently replaced by the
commutativity formula (1.4) or (2.2.7).
In Section 3 we give an analogue of the notion of endomorphism ring of a vector
space by means of local systems of vertex operators.
In Section 4 we study vertex operator superalgebras and modules
associated to some well-known infinite-dimensional Lie algebras or Lie
superalgebras. In Section 5
we give the semisimple representation theory for
vertex operator algebras associated to standard modules
for an affine Lie algebra. In Section 6 we give an analogue of the
notion of the space of linear homomorphisms.

\vspace{0.5cm}
\noindent {\large \bf Acknowledgement}\\
This paper was greatly influenced by Professor James Lepowsky's lectures on
vertex operator algebras at Rutgers for the last two years.
We would like to thank Professor Lepowsky for much
insightful advice, many stimulating discussions and patient help in
both mathematics and paper writing. We are grateful to
Professor Chongying Dong from whom I have learned a lot about vertex
operator algebra theory, in particular, the proof of
Proposition 3.2.7. We believe that some results presented here are
known to him. We would like also to thank Professor Robert Wilson for
reading this paper. Four
years ago, Professor Wilson had obtained some results (unpublished) about an
analogue of endomorphism ring for vertex operator algebras; the
present results are in a different direction.

\makeatletter
\@addtoreset{equation}{subsection}

\makeatother
\makeatletter

\section{Vertex superalgebras and modules}
In this section, we shall prove that the Jacobi identity for vertex
superalgebra can be equivalently replaced by the supercommutativity
formula (1.4) (and (2.2.7)) or by the skew symmetry (2.2.5) and
the associativity (2.2.9). The same proof shows that
the Jacobi identity for module of a vertex superalgebra can be
replaced by the associativity (2.2.9). This proves that Borcherds' notion
of vertex algebra [B] is essentially equivalent to the notion of
vertex operator algebra formulated by
Frenkel, Lepowsky and Meurman [FLM]. (This is presumably known to Borcherds.)
Since  our approach is ``matrix-coefficient''-free, it meets our
needs for dealing with arbitrary vertex superalgebras for which there
may be no grading at all.
All the results of this section could be easily extended to ``colored vertex
operator superalgebra'' (cf. [Xu]).

\subsection{Formal calculus}
In this subsection we shall present some  elementary operations and
properties of
formal series. Our notations agree with those in [FLM] and [FHL].
Let $x$, $y$, $z$, $z_{0}$, $z_{1},\cdots$ be
commuting formal variables. For a vector space $V$, we set
\begin{eqnarray}
V\{z\}=\{\sum_{n\in {\bf C}}v_{n}z^{n}|v_{n}\in V\},\;
V[[z,z^{-1}]]=\{\sum_{n\in {\bf Z}}v_{n}z^{n}|v_{n}\in V\}.
\end{eqnarray}
We set the following subspaces of $V[[z,z^{-1}]]$:
\begin{eqnarray}
& &V[z]=\{\sum_{n\in {\bf N}}v_{n}z^{n}|v_{n}\in V,v_{n}=0\;\mbox{ for }
n\mbox{ suficiently large}\},\\
& &V[z,z^{-1}]=\{\sum_{n\in {\bf Z}}v_{n}z^{n}|v_{n}\in V,v_{n}=0\;\mbox{ for
all but finitely many }n\},\\
& &V[[z]]=\{\sum_{n\in{\bf N}}v_{n}z^{n}|v_{n}\in V\},\\
& &V((z))=\{\sum_{n\in {\bf Z}}v_{n}z^{n}|v_{n}\in V,v_{n}=0\;\mbox{
for }n\;\mbox{suficiently small}\}.
\end{eqnarray}
For $f(z)=\sum_{n\in{\bf Z}}v_{n}z^{n}\in V[[z,z^{-1}]]$, the formal
derivative is defined to be
\begin{eqnarray}
{d\over dz}f(z)=f'(z)=\sum_{n\in {\bf Z}}nv_{n}z^{n-1},
\end{eqnarray}
and the formal residue is defined as follows:
\begin{eqnarray}
{\rm Res}_{z}f(z)=v_{-1}\;\;\;\mbox{(the coefficient of $z^{-1}$ in $f(z)$)}.
\end{eqnarray}
If $f(z)\in {\bf C}((z)), g(z)\in V((z))$, we have:
\begin{eqnarray}
{\rm Res}_{z}(f'(z)g(z))=-{\rm Res}_{z}(f(z)g'(z)).
\end{eqnarray}
For $\alpha\in {\bf C}$, as a formal series,
$(z_{1}+z_{2})^{\alpha}$ is defined to be
\begin{eqnarray}
(z_{1}+z_{2})^{\alpha}=\sum_{k=0}^{\infty}\left(\begin{array}{c}\alpha\\k
\end{array}\right)z_{1}^{\alpha -k}z_{2}^{k},
\end{eqnarray}
where
$\displaystyle{\left(\begin{array}{c}\alpha\\k\end{array}\right)=\frac{\alpha
(\alpha-1)\cdots (\alpha-k+1)}{k!}}$.
If $f(z)\in V\{z\}$, then we have the following Taylor formula:
\begin{eqnarray}
e^{z_{0}{\partial\over\partial z}}f(z)=f(z+z_{0}).
\end{eqnarray}

In vertex algebra theory, one of the most important formal series is the
formal $\delta$-function, which is defined to be
\begin{eqnarray}
\delta(z)=\sum_{n\in {\bf Z}}z^{n}\in {\bf C}[[z,z^{-1}]].
\end{eqnarray}
Thus
\begin{eqnarray}
\delta\left(\frac{z_{1}-z_{2}}{z_{0}}\right)=\sum_{n\in {\bf Z}}z_{0}^{-n}
(z_{1}-z_{2})^{n}=\sum_{n\in {\bf Z}}\sum_{k\in {\bf N}}(-1)^{k}\left(
\begin{array}{c}n\\k\end{array}\right)z_{0}^{-n}z_{1}^{n-k}z_{2}^{k}.
\end{eqnarray}
Furthermore, we have:
\begin{eqnarray}
& &z_{0}^{-1}\delta\left(\frac{z_{1}-z_{2}}{z_{0}}\right)-z_{0}^{-1}\delta
\left(\frac{z_{2}-z_{1}}{-z_{0}}\right)=z_{2}^{-1}\delta\left(\frac{z_{1}
-z_{0}}{z_{2}}\right),\\
& &z_{0}^{-1}\delta\left(\frac{z_{1}-z_{2}}{z_{0}}\right)=z_{1}^{-1}\delta
\left(\frac{z_{0}+z_{2}}{z_{1}}\right).
\end{eqnarray}

{\bf Lemma 2.1.1.} {\it As formal series, we have}
\begin{eqnarray}
{\partial\over\partial z_{0}}\left(z_{0}^{-1}\delta\left(\frac{z_{1}-z_{2}}
{z_{0}}\right)\right)={\partial\over\partial z_{2}}\left(z_{0}^{-1}\delta
\left(\frac{z_{1}-z_{2}}{z_{0}}\right)\right)=-{\partial\over\partial z_{1}}
\left(z_{0}^{-1}\delta\left(\frac{z_{1}-z_{2}}{z_{0}}\right)\right).
\end{eqnarray}

{\sl Proof.} It is clear that $\displaystyle{{\partial\over\partial
z_{2}}\left(z_{0}^{-1}
\delta\left(\frac{z_{1}-z_{2}}{z_{0}}\right)\right)=-{\partial\over\partial
z_{1}}\left(z_{0}^{-1}\delta\left(\frac{z_{1}-z_{2}}{z_{0}}\right)\right)}$.
By (2.1.14), we have
\begin{eqnarray}
{\partial\over\partial z_{0}}\left(z_{0}^{-1}\delta\left(\frac{z_{1}-z_{2}}
{z_{0}}\right)\right)&=&{\partial\over\partial z_{0}}\left(z_{1}^{-1}\delta
\left(\frac{z_{0}+z_{2}}{z_{1}}\right)\right)\nonumber\\
&=&{\partial\over\partial z_{2}}\left(z_{1}^{-1}\delta
\left(\frac{z_{0}+z_{2}}{z_{1}}\right)\right)\nonumber\\
&=&{\partial\over\partial z_{2}}\left(z_{0}^{-1}\delta
\left(\frac{z_{1}-z_{2}}{z_{0}}\right)\right).\;\;\;\;\Box
\end{eqnarray}

{\bf Lemma 2.1.2 [FLM].} a) {\it If} $f(z_{1},z_{2})\in
V[[z_{1},z_{1}^{-1}]]((z_{2}))$, {\it then}
\begin{eqnarray}
\delta\left(\frac{z_{0}+z_{2}}{z_{1}}\right)f(z_{1},z_{2})=
\delta\left(\frac{z_{0}+z_{2}}{z_{1}}\right)f(z_{0}+z_{2},z_{2}).
\end{eqnarray}
b) {\it If} $f(z_{0},z_{2})\in
V[[z_{0},z_{0}^{-1}]][[z_{2}]]$, {\it then}
\begin{eqnarray}
\delta\left(\frac{z_{1}-z_{2}}{z_{0}}\right)f(z_{0},z_{2})=
\delta\left(\frac{z_{1}-z_{2}}{z_{0}}\right)f(z_{0},z_{1}-z_{0}).
\end{eqnarray}

{\bf Lemma 2.1.3.} {\it If} $m,n\in {\bf N}, m>n$, {\it then}
\begin{eqnarray}
(z_{1}-z_{2})^{m}\delta ^{(n)}\left({z_{1}\over z_{2}}\right)=0.
\end{eqnarray}

{\sl Proof.} We prove this lemma by using induction on $n$. If $n=0$,
it follows from Lemma 2.1.2 that
\begin{eqnarray}
(z_{1}-z_{2})^{m}\delta\left({z_{1}\over z_{2}}\right)=0\;\;\;\mbox{
for any }0<m\in {\bf Z}.
\end{eqnarray}
Assume that
$\displaystyle{(z_{1}-z_{2})^{k}\delta^{(n)}\left({z_{1}\over z_{2}}\right)=0}$
for any
$k>n$. If $m>n+1$, then by assumption,
\begin{eqnarray}
(z_{1}-z_{2})^{m-1}\delta^{(n)}\left({z_{1}\over z_{2}}\right)=0,\;
(z_{1}-z_{2})^{m}\delta^{(n)}\left({z_{1}\over z_{2}}\right)=0.
\end{eqnarray}
Differentiating $\displaystyle {(z_{1}-z_{2})^{m}\delta^{(n)}\left({z_{1}\over
z_{2}}\right)=0}$ with respect to $z_{1}$, then using (2.1.21) we obtain
\begin{eqnarray}
(z_{1}-z_{2})^{m}\delta^{(n+1)}\left({z_{1}\over z_{2}}\right)=0.
\end{eqnarray}
This finishes the induction procedure. $\;\;\;\;\Box$

{\bf Lemma 2.1.4.} {\it Let $V$ be any vector space and let
$f_{0}(z_{2}),\cdots,f_{m}(z_{2})\in V[[z_{2},z_{2}^{-1}]]$. Then}
\begin{eqnarray}
z_{1}^{-1}\delta\left(\frac{z_{2}}{z_{1}}\right)f_{0}(z_{2})+\cdots +
z_{1}^{-m-1}\delta^{(m)}\left(\frac{z_{2}}{z_{1}}\right)f_{m}(z_{2})=0
\end{eqnarray}
{\it if and only if $f_{k}(z_{2})=0$ for $k=0,\cdots,m$.}

{\sl Proof.} We shall prove this lemma by contradiction. Suppose
$f_{k}(z_{2})\ne 0$ for some $k$. Let $n$ be the nonnegative integer
such that $f_{n}(z_{2})\ne 0$ and $f_{k}(z_{2})=0$ for $n<k\le m$.
Applying $\displaystyle{{\rm Res}_{z_{1}}z_{1}^{n+1}(z_{1}-z_{2})^{n}}$ to the
left hand side of (2.1.23), by Lemma 2.1.3 we obtain
\begin{eqnarray}
0&=&{\rm Res}_{z_{1}}(z_{1}-z_{2})^{n}\delta^{(n)}\left(
\frac{z_{2}}{z_{1}}\right)f_{n}(z_{2})\nonumber\\
&=&{\rm Res}_{z_{1}}(z_{1}-z_{2})^{n}\left({\partial\over\partial
z_{1}}\right)^{n} \delta\left(
\frac{z_{1}}{z_{2}}\right)z_{2}^{n}f_{n}(z_{2})\nonumber\\
&=&(-1)^{n}n!z_{2}^{n+1}f_{n}(z_{2}).
\end{eqnarray}
Therefore $f_{n}(z_{2})=0$. This is a contradiction. $\;\;\;\;\Box$

\newpage
\subsection{Vertex (operator) superalgebras}

Let $M=M^{0}\oplus M^{1}$ be any $({\bf Z}/2{\bf Z})$-graded vector
space. Then any element $u$ in $M^{0}$ (resp. $M^{1}$) is said to be
{\it even} (resp. {\it odd}).
For any homogeneous element $u$, we define $|u|=0$ if $u$
is even, $|u|=1$ if $u$ is odd. If $M$ and $W$ are any two $({\bf
Z}/2{\bf Z})$-graded vector spaces, we define
$\varepsilon_{u,v}=(-1)^{|u||v|}$ for any homogeneous elements $u\in
M, v\in W$.

The following definition of vertex (operator) superalgebra (cf. [T])
is formulated according to FLM's definition of vertex operator algebra ([FLM],
[FHL]) and Borcherds' definition of vertex algebra [B].

{\bf Definition 2.2.1. } A {\it vertex superalgebra} is a quadruple
$(V,D,{\bf 1},Y)$, where $V=V^{0}\oplus V^{1}$ is a $({\bf
Z}/2{\bf Z})$-graded vector space, $D$ is an endomorphism
of $V$, ${\bf 1}$ is a specified vector called the {\it vacuum} of $V$,
and $Y$ is a linear map
\begin{eqnarray}
Y(\cdot,z):& &V\rightarrow ({\rm End}V)[[z,z^{-1}]];\nonumber\\
& &a\mapsto Y(a,z)=\sum_{n\in{\bf Z}}a_{n}z^{-n-1}\;\;(\mbox{where }
a_{n}\in {\rm End}V)
\end{eqnarray}
such that
\begin{eqnarray*}
(V1)& &\mbox{For any }a,b\in V, a_{n}b=0\;\;\;\mbox{ for }n
\mbox{ sufficiently large;}\\
(V2)& &[D,Y(a,z)]=Y(D(a),z)={d\over dz}Y(a,z)\;\;\mbox{  for any }a\in V;\\
(V3)& &Y({\bf 1},z)={\rm id}_{V}\;\;\;\mbox{(the identity operator of $V$)};\\
(V4)& &Y(a,z){\bf 1}\in ({\rm End}V)[[z]] \mbox{ and }\lim_{z \rightarrow
0}Y(a,z){\bf 1}=a\;\;\mbox{  for any }a\in V;\\
(V5)& &\mbox{For $({\bf Z}/2{\bf Z})$-homogeneous elements }a,b\in V,
\mbox{ the following {\it Jacobi identity} holds:}
\end{eqnarray*}
\begin{eqnarray}
& &z_{0}^{-1}\delta\left(\frac{z_{1}-z_{2}}{z_{0}}\right)Y(a,z_{1})Y(b,z_{2})
-\varepsilon_{a,b}z_{0}^{-1}\delta\left(\frac{z_{2}-z_{1}}{-z_{0}}\right)
Y(b,z_{2})Y(a,z_{1})\nonumber \\
&=&z_{2}^{-1}\delta\left(\frac{z_{1}-z_{0}}{z_{2}}\right)Y(Y(a,z_{0})b,z_{2}).
\end{eqnarray}

This completes the definition of vertex superalgebra. The following
are consequences:
\begin{eqnarray}
& &Y(a,z){\bf 1}=e^{zD}a\;\;\;\mbox{ for any }a\in V,\\
& &e^{z_{0}D}Y(a,z)e^{-z_{0}D}=Y(a,z+z_{0})\;\;\;\mbox{ for any }a\in V,\\
& &Y(a,z)b=\varepsilon_{a,b}e^{zD}Y(b,-z)b\;\;\;\mbox{ for $({\bf
Z}/2{\bf Z})$-homogeneous }a, b\in V.
\end{eqnarray}

For any nonnegative integer $k$, taking ${\rm Res}_{z_{0}}z_{0}^{k}$
of the Jacobi identity, we obtain:
\begin{eqnarray}
& &(z_{1}-z_{2})^{k}[Y(a,z_{1}),Y(b,z_{2})]_{\pm}:=\nonumber\\
&=&(z_{1}-z_{2})^{k}\left(Y(a,z_{1})Y(b,z_{2})-\varepsilon_{a,b}
Y(b,z_{2})Y(a,z_{1})\right)\nonumber\\
&=&{\rm Res}_{z_{0}}z_{2}^{-1}\delta\left(
\frac{z_{1}-z_{0}}{z_{2}}\right)z_{0}^{k}Y(Y(a,z_{0})b,z_{2})\nonumber \\
&=&\sum_{i=0}^{\infty}\frac{1}{i!}\left(\left({\partial\over\partial z_{2}}
\right)^{i}z_{1}^{-1}\delta\left(\frac{z_{2}}{z_{1}}\right)\right)
Y(a_{k+i}b,z_{2}).
\end{eqnarray}
If $k=0$, we get the {\it supercommutator formula}.
Let $m$ be a positive integer such that
$a_{n}b=0$ for $n\ge m$. Then we obtain the following {\it
supercommutativity}:
\begin{eqnarray}
(z_{1}-z_{2})^{m}Y(a,z_{1})Y(b,z_{2})=
\varepsilon_{a,b} (z_{1}-z_{2})^{m}Y(b,z_{2})Y(a,z_{1}).
\end{eqnarray}

Taking ${\rm
Res}_{z_{1}}$ of the Jacobi identity, we obtain the following {\it
iterate formula}:
\begin{eqnarray}
& &Y(Y(a,z_{0})b,z_{2})\nonumber \\
&=&{\rm Res}_{z_{1}}\left(z_{0}^{-1}\delta\left(\frac{z_{1}-z_{2}}{z_{0}}
\right)Y(a,z_{1})Y(b,z_{2})-\varepsilon_{a,b}
z_{0}^{-1}\delta\left(\frac{-z_{2}+z_{1}}{z_{0}}
\right)Y(b,z_{2})Y(a,z_{1})\right)\nonumber \\
&=&Y(a,z_{0}+z_{2})Y(b,z_{2})-\varepsilon_{a,b}Y(b,z_{2})(Y(a,z_{0}+z_{2})-Y(a,z_{2}+z_{0})).
\end{eqnarray}
For any $c\in V$, let $m$ be a positive integer such that
$z^{m}Y(a,z)c$ involves only positive powers of $z$. Then we obtain
 the following {\it associativity}:
\begin{eqnarray}
(z_{0}+z_{2})^{m}Y(Y(a,z_{0})b,z_{2})c=(z_{0}+z_{2})^{m}Y(a,z_{0}+z_{2})
Y(b,z_{2})c.
\end{eqnarray}

A vertex superalgebra $V$ is called a {\it vertex operator
superalgebra} (cf. [T]) if
there is another distinguished vector $\omega$ of $V$ such that
\begin{eqnarray*}
(V6)& &[L(m),L(n)]=(m-n)L(m+n)+\frac{m^{3}-m}{12}\delta_{m+n,0}({\rm rank}V)\\
& &\mbox{for }m,n\in {\bf Z},\mbox{ where }Y(\omega,z)=\sum_{n\in {\bf
Z}}L(n)z^{-n-2},\;{\rm
rank}V\in {\bf C};\\
(V7)& &L(-1)=D,\; i.e., Y(L(-1)a,z)={d\over dz}Y(a,z)\;\;\;\mbox{ for any
}a\in V;\\
(V8)& &V\;\mbox{ is }{1\over 2}{\bf Z}\mbox{-graded such that
}V=\oplus_{n\in {1\over 2}{\bf
Z}}V_{(n)},\; L(0)|_{V_{(n)}}=n{\rm id},\;\dim V_{(n)} <\infty,\\
& &\mbox{ and } V_{(n)}=0\;\;\mbox{ for }n\mbox{ sufficiently small. }
\end{eqnarray*}

{\bf Remark 2.2.2.} In the definition of vertex operator
superalgebra, since the first equality of (V2) follows from
the Jacobi identity, it is enough only to assume the second equality.
But for vertex superalgebra, it is necessary to assume both of them.

{\bf Remark 2.2.3.}  Let $M$ be any vector space with a set
$\{ L(m)|m\in {\bf Z}\}$ of endomorphisms of $M$ and let
 $\ell$ be a complex number. Set $L(z)=\sum_{n\in {\bf Z}}L(n)z^{-n-2}$.
Then the Virasoro relation (V6) with central charge $\ell$ is equivalent
to the following equation:
\begin{eqnarray}
& &[L(z_{1}),L(z_{2})]\nonumber \\
&=&\sum_{m,n\in {\bf Z}}[L(m),L(n)]z_{1}^{-m-2}z_{2}^{-n-2}\nonumber\\
&=&\sum_{m,n\in {\bf Z}}\left((m-n)L(m+n)+{m^{3}-m\over
12}\delta_{m+n,0}\ell\right) z_{1}^{-m-2}z_{2}^{-n-2}\nonumber\\
&=&\sum_{m,n\in {\bf
Z}}\left((-m-n-2)L(m+n)+2(m+1)L(m+n)\right)z_{1}^{-m-2}z_{2}^{-n-2}\nonumber \\
& &+\ell\sum_{m\in {\bf Z}}\frac{m^{3}-m}{12}z_{1}^{-m-2}z_{2}^{m-2}\nonumber\\
&=&z_{1}^{-1}\delta\left({z_{2}\over z_{1}}\right)L'(z_{2})+2z_{1}^{-2}
\delta\left({z_{2}\over z_{1}}\right)L(z_{2})+{\ell \over
12}z_{1}^{-4}\delta ^{(3)}\left({z_{2}\over z_{1}}\right).
\end{eqnarray}
Then it follows from Lemma 2.1.4 and the commutator formula (2.2.6)
that the Virasoro relation (V6) and the $L(-1)$-derivative relation
(V7) are equivalent to the following conditions:
\begin{eqnarray}
& &Y(\omega_{0}a,z)={d\over dz}Y(a,z)\;\;\;\mbox{for any }a\in V,\\
& &\omega_{1}\omega =2\omega,\;
\omega_{2}\omega =0,\;\omega_{3}\omega ={\ell\over 2}{\bf 1},\;
\omega_{n}\omega=0\;\;\;\mbox{for }n>3.
\end{eqnarray}

{\bf Proposition 2.2.4}. {\it In the definition of vertex superalgebra, the
Jacobi identity can be equivalently substituted by the
supercommutativity formula (2.2.7).}

{\bf Remark 2.2.5}. Notice that the supercommutativity formula (2.2.7) is
really a supercommutativity formula for ``left multiplications.'' Let
us give a ``proof'' of Proposition 2.2.4 at the
classical level as follows: Let
$A$ be any algebra with a right identity 1 and denote by $\ell _{a}$ the left
multiplication by an element $a$. Suppose that
$\ell_{a}\ell_{b}=\ell_{b}\ell_{a}$ for any $a,b\in A$. Then
 $a(bc)=b(ac)$ for any $a,b,c\in A$. Setting
$c=1$, we obtain the commutativity $ab=ba$. Furthermore, we obtain the
associativity:
\begin{eqnarray}
a(cb)=a(bc)=b(ac)=(ac)b\;\;\;\mbox{ for any }a,b,c\in A.
\end{eqnarray}
Therefore $A$ is a commutative associative algebra.

{\sl Proof of Proposition 2.2.4}. Our proof, which consists of three steps,
is exactly an analogue of the argument given in Remark 2.2.5.

(1) The {\it skew-symmetry} (2.2.5) holds. Let $k$ be a positive
integer such that
$b_{m}a=0$ for all $m\ge k$ and that the supercommutativity formula (2.2.7)
holds. Then
\begin{eqnarray}
& &(z_{1}-z_{2})^{k}Y(a,z_{1})Y(b,z_{2}){\bf 1}\nonumber\\
&=&\varepsilon_{a,b}(z_{1}-z_{2})^{k}Y(b,z_{2})Y(a,z_{1}){\bf 1}\nonumber\\
&=&\varepsilon_{a,b}(z_{1}-z_{2})^{k}Y(b,z_{2})e^{z_{1}D}a\nonumber \\
&=&\varepsilon_{a,b}(z_{1}-z_{2})^{k}e^{z_{1}D}Y(b,z_{2}-z_{1})a.
\end{eqnarray}
Since $(z_{1}-z_{2})^{k}Y(b,z_{2}-z_{1})a$ involves only nonnegative
powers of $(z_{2}-z_{1})$, we may set
$z_{2}=0$. Thus
\begin{eqnarray}
z_{1}^{k}Y(a,z_{1})b=\varepsilon_{a,b}z_{1}^{k}e^{z_{1}D}Y(b,-z_{1})a.
\end{eqnarray}
Multiplying both sides of (2.2.15) by $z_{1}^{-k}$ we obtain
 $Y(a,z_{1})b=\varepsilon_{a,b}e^{z_{1}D}Y(b,-z_{1})a$.

(2) The associativity formula (2.2.9) holds.
For any $({\bf Z}/2{\bf Z})$-homogeneous $a,c\in V$, let $k$ be a
positive integer such that
the supercommutativity formula (2.2.7) for $(a,c)$ holds. Then for any
$b\in V$, we have:
\begin{eqnarray}
& &(z_{0}+z_{2})^{k}Y(a,z_{0}+z_{2})Y(b,z_{2})c\nonumber\\
&=&\varepsilon_{b,c}(z_{0}+z_{2})^{k}Y(a,z_{0}+z_{2})e^{z_{2}D}Y(c,-z_{2})b
\nonumber\\
&=&\varepsilon_{b,c}e^{z_{2}D}(z_{0}+z_{2})^{k}Y(a,z_{0})
Y(c,-z_{2})b\nonumber\\
&=&\varepsilon_{b,c}\varepsilon_{a,c}e^{z_{2}D}(z_{0}+z_{2})^{k}Y(c,-z_{2})
Y(a,z_{0})b\nonumber\\
&=&(z_{0}+z_{2})^{k}Y(Y(a,z_{0})b,z_{2})c.\end{eqnarray}

(3) The Jacobi identity holds. Choosing $k$ in (2) such that $a_{m}c=0$ for all
$m\ge k$, we get
\begin{eqnarray}
& &z_{0}^{k}z_{1}^{k}\left(z_{0}^{-1}\delta\left(\frac{z_{1}-z_{2}}{z_{0}}
\right)Y(a,z_{1})Y(b,z_{2})c-\varepsilon_{a,b}z_{0}^{-1}\delta\left(
\frac{-z_{2}+z_{1}}
{z_{0}}\right)Y(b,z_{2})Y(a,z_{1})c\right)\nonumber\\
&=&z_{0}^{-1}\delta\left(\frac{z_{1}-z_{2}}{z_{0}}\right)z_{1}^{k}
(z_{1}-z_{2})^{k}Y(a,z_{1})Y(b,z_{2})c\nonumber\\
& &-\varepsilon_{a,b}z_{0}^{-1}\delta\left(\frac{-z_{2}+z_{1}}{z_{0}}\right)
z_{1}^{k}(z_{1}-z_{2})^{k}Y(b,z_{2})Y(a,z_{1})c\nonumber\\
&=&z_{2}^{-1}\delta\left(\frac{z_{1}-z_{0}}{z_{2}}\right)\varepsilon_{a,b}
\left(z_{1}^{k}(z_{1}-z_{2})^{k}Y(b,z_{2})Y(a,z_{1})c\right)\nonumber\\
&=&z_{2}^{-1}\delta\left(\frac{z_{1}-z_{0}}{z_{2}}\right)\varepsilon_{a,b}
\left(z_{0}^{k}(z_{0}+z_{2})^{k}Y(b,z_{2})Y(a,z_{2}+z_{0})c\right).
\end{eqnarray}
Since $a_{m}c=0$ for all $m\ge k$,
$(z_{0}+z_{2})^{k}Y(a,z_{2}+z_{0})c$ involves only nonnegative powers
of $(z_{2}+z_{0})$, so that
\begin{eqnarray}
z_{0}^{k}(z_{0}+z_{2})^{k}Y(b,z_{2})Y(a,z_{2}+z_{0})c
=z_{0}^{k}(z_{0}+z_{2})^{k}Y(b,z_{2})Y(a,z_{0}+z_{2})c.
\end{eqnarray}
Therefore
\begin{eqnarray}
& &z_{0}^{k}z_{1}^{k}\left(z_{0}^{-1}\delta\left(\frac{z_{1}-z_{2}}{z_{0}}
\right)Y(a,z_{1})Y(b,z_{2})c-\varepsilon_{a,b}z_{0}^{-1}\delta\left(
\frac{z_{2}-z_{1}}
{-z_{0}}\right)Y(b,z_{2})Y(a,z_{1})c\right)\nonumber\\
&=&z_{2}^{-1}\delta\left(\frac{z_{1}-z_{0}}{z_{2}}\right)\varepsilon_{a,b}
\left(z_{0}^{k}(z_{0}+z_{2})^{k}Y(b,z_{2})Y(a,z_{0}+z_{2})c\right)
\nonumber\\
&=&z_{2}^{-1}\delta\left(\frac{z_{1}-z_{0}}{z_{2}}\right)\left(z_{0}^{k}
(z_{0}+z_{2})^{k}Y(a,z_{0}+z_{2})Y(b,z_{2})c\right)\nonumber\\
&=&z_{2}^{-1}\delta\left(\frac{z_{1}-z_{0}}{z_{2}}\right)\left(z_{0}^{k}
(z_{0}+z_{2})^{k}Y(Y(a,z_{0})b,z_{2})c\right)\nonumber\\
&=&z_{2}^{-1}\delta\left(\frac{z_{1}-z_{0}}{z_{2}}\right)\left(z_{0}^{k}
z_{1}^{k}Y(Y(a,z_{0})b,z_{2})c\right)\nonumber\\
&=&z_{0}^{k}z_{1}^{k}z_{2}^{-1}\delta\left(\frac{z_{1}-z_{0}}{z_{2}}\right)
Y(Y(a,z_{0})b,z_{2})c.
\end{eqnarray}
Then the Jacobi identity follows. $\;\;\;\;\Box$

{\bf Proposition 2.2.6.} {\it The Jacobi identity for vertex
superalgebra $V$ can be equivalently replaced by the skew symmetry
(2.2.5) and the associativity
formula (2.2.9).}

{\sl Proof}.  For any
$a,b,c\in V$, let $k$ be a positive integer such that $z^{k}Y(b,z)c$
involves only positive powers of $z$ and that the following
associativities hold:
\begin{eqnarray*}(z_{0}+z_{2})^{k}Y(a,z_{0}+z_{2})Y(b,z_{2})c
&=&(z_{0}+z_{2})^{k}Y(Y(a,z_{0})b,z_{2})c,\\
(-z_{0}+z_{1})^{k}Y(b,-z_{0}+z_{1})Y(a,z_{1})c&=&(-z_{0}+z_{1})^{k}
Y(Y(b,-z_{0})a,z_{1})c.
\end{eqnarray*}
Then
\begin{eqnarray*}
& &z_{1}^{k}z_{2}^{k}\left(z_{0}^{-1}\delta\left(\frac{z_{1}-z_{2}}
{z_{0}}\right)Y(a,z_{1})Y(b,z_{2})c-\varepsilon_{a,b}z_{0}^{-1}\delta\left(
\frac{-z_{2}+z_{1}}
{z_{0}}\right)Y(b,z_{2})Y(a,z_{1})c\right)\\
&=&z_{0}^{-1}\delta\left(\frac{z_{1}-z_{2}}
{z_{0}}\right)\left((z_{0}+z_{2})^{k}z_{2}^{k}Y(Y(a,z_{0})b,z_{2})c\right)\\
& &-\varepsilon_{a,b}z_{0}^{-1}\delta\left(\frac{-z_{2}+z_{1}}
{z_{0}}\right)\left(z_{1}^{k}(-z_{0}+z_{1})^{k}Y(Y(b,-z_{0})a,z_{1})c\right)\\
&=&z_{0}^{-1}\delta\left(\frac{z_{1}-z_{2}}
{z_{0}}\right)\left((z_{0}+z_{2})^{k}z_{2}^{k}Y(Y(a,z_{0})b,z_{2})c\right)\\
& &-\varepsilon_{a,b}z_{0}^{-1}\delta\left(\frac{-z_{2}+z_{1}}
{z_{0}}\right)\left(z_{1}^{k}(-z_{0}+z_{1})^{k}Y(e^{-z_{0}D}Y(a,z_{0})
b,z_{1})c\right)\\
&=&z_{0}^{-1}\delta\left(\frac{z_{1}-z_{2}}
{z_{0}}\right)\left((z_{0}+z_{2})^{k}z_{2}^{k}Y(Y(a,z_{0})b,z_{2})c\right)\\
& &-\varepsilon_{a,b}z_{0}^{-1}\delta\left(\frac{-z_{2}+z_{1}}
{z_{0}}\right)\left(z_{1}^{k}(-z_{0}+z_{1})^{k}Y(Y(a,z_{0})b,z_{1}-z_{0})c
\right).\end{eqnarray*}
Since $z_{2}^{k}(z_{0}+z_{2})^{k}Y(Y(a,z_{0})b,z_{2})c=(z_{0}+z_{2})^{k}
Y(a,z_{0}+z_{2})(z_{2}^{k}Y(b,z_{2})c)$ involves only
positive powers of $z_{2}$, by (2.1.18) we have:
\begin{eqnarray}
& &z_{0}^{-1}\delta\left(\frac{z_{1}-z_{2}}
{z_{0}}\right)\left((z_{0}+z_{2})^{k}z_{2}^{k}Y(Y(a,z_{0})b,z_{2})c\right)
\nonumber\\
&=&z_{0}^{-1}\delta\left(\frac{z_{1}-z_{2}}
{z_{0}}\right)\left(z_{1}^{k}(z_{1}-z_{0})^{k}Y(Y(a,z_{0})b,z_{1}-z_{0})c
\right).
\end{eqnarray}
Thus
\begin{eqnarray}
& &z_{1}^{k}z_{2}^{k}\left(z_{0}^{-1}\delta\left(\frac{z_{1}-z_{2}}
{z_{0}}\right)Y(a,z_{1})Y(b,z_{2})c-\varepsilon_{a,b}z_{0}^{-1}\delta\left(
\frac{-z_{2}+z_{1}}
{z_{0}}\right)Y(b,z_{2})Y(a,z_{1})c\right)\nonumber\\
&=&z_{0}^{-1}\delta\left(\frac{z_{1}-z_{2}}
{z_{0}}\right)\left(z_{1}^{k}(z_{1}-z_{0})^{k}Y(Y(a,z_{0})b,z_{1}-z_{0})c
\right)\nonumber\\
& &-\varepsilon_{a,b}z_{0}^{-1}\delta\left(\frac{-z_{2}+z_{1}}
{z_{0}}\right)\left(z_{1}^{k}(-z_{0}+z_{1})^{k}Y(Y(a,z_{0})b,z_{1}-z_{0})c
\right)\nonumber\\
&=&z_{2}^{-1}\delta\left(\frac{z_{1}-z_{0}}
{z_{2}}\right)\left(z_{1}^{k}(z_{1}-z_{0})^{k}Y(Y(a,z_{0})b,z_{1}-z_{0})c
\right)\nonumber\\
&=&z_{1}^{k}z_{2}^{k}z_{2}^{-1}\delta\left(\frac{z_{1}-z_{0}}{z_{2}}\right)
Y(Y(a,z_{0})b,z_{2})c.
\end{eqnarray}
Multiplying both sides by $z_{1}^{-k}z_{2}^{-k}$, we obtain the Jacobi
identity.$\;\;\;\;\Box$

In [B], Borcherds first defined the notion of vertex algebra with a
set of axioms consisting of (V1), (V3), (V4), the skew-symmetry
(2.2.5) and the iterate formula (2.2.9). In the skew-symmetry (2.2.5),
the operator $D$ is defined by $Da=a_{-2}{\bf 1}$ for $a\in V$. It
can be easily proved that (V2) follows from the other axioms of Borcherds'.
Therefore, we have:

{\bf Corollary 2.2.7}. {\it Borcherds' definition [B] and Definition 2.1
for a vertex superalgebra are equivalent.}

\subsection{Modules for vertex superalgebras}

\indent
{\bf Definition 2.3.1}. Let $(V,{\bf 1},D,Y)$ be a vertex superalgebra. A
$V$-module is a triple $(M, d, Y_{M})$ where $M$ is a $({\bf Z}/2{\bf
Z})$-graded vector space, $d$ is an endomorphism of $M$ and $Y_{M}$ is
a linear map
$Y_{M}(\cdot,z):\;V\rightarrow ({\rm End}M)[[z,z^{-1}]];
\;a\mapsto Y_{M}(a,z)=\sum_{n\in {\bf
Z}}a_{n}z^{-n-1}$ (where $a_{n}\in {\rm End}M)$ satisfying the
following conditions:
\begin{eqnarray*}
(M1)& &\mbox{For any }a\in V,u\in M, a_{n}u=0\;\;\;\mbox{ for }n
\mbox{ sufficient large};\\
(M2)& &Y_{M}({\bf 1},z)={\rm id}_{M};\\
(M3)& &[d,Y_{M}(a,z)]=Y_{M}(D(a),z)={d\over dz}Y_{M}(a,z)\;\;\mbox{
for any }a\in V;\\
(M4)& &\mbox{For $({\bf Z}/2{\bf Z})$-homogeneous }a,b\in V,\mbox{ the
following Jacobi identity holds:}
\end{eqnarray*}
\begin{eqnarray}
& &z_{0}^{-1}\delta\left(\frac{z_{1}-z_{2}}{z_{0}}\right)Y_{M}(a,z_{1})
Y_{M}(b,z_{2})-\varepsilon_{a,b}z_{0}^{-1}\delta\left(\frac{z_{2}-z_{1}}
{-z_{0}}\right)Y_{M}(b,z_{2})Y_{M}(a,z_{1})\nonumber\\
&=&z_{2}^{-1}\delta\left(\frac{z_{1}-z_{0}}{z_{2}}\right)
Y_{M}(Y(a,z_{0})b,z_{2}).
\end{eqnarray}

This completes the definition of module for a vertex superalgebra. A
$V$-module $M$ is called a {\it faithful} module if $Y_{M}(\cdot,z)$
is injective.
If $V$ is a vertex operator superalgebra, a module $M$ for $V$ being a vertex
superalgebra is called a {\it weak} module for $V$ being a VOSA.
 A weak $V$-module $M$ is
said to be {\it $\frac{1}{2}{\bf Z}$-graded} if $M=\oplus _{n\in
{1\over 2}{\bf Z}}M(n)$ such that
\begin{eqnarray*}
(M5)& &a_{m}M(n)\subseteq M(n+ r-m-1)\;\;\mbox{ for }a\in V_{(r)}, m\in
{\bf Z},n, r\in \frac{1}{2}{\bf Z}.
\end{eqnarray*}
Furthermore, a weak $V$-module $M$
is called a {\it generalized module} [HL] for $V$ being a VOSA if
$M=\oplus _{\alpha\in {\bf C}}M_{(\alpha )}$ such that

$(M6)\;\;\;\;L(0)u=\alpha u\;\;\;\mbox{for }\alpha \in {\bf C},u\in
M_{(\alpha)},$\\
and a generalized module $M$ is called a {\it module} for $V$ being a VOSA if

$(M7)\;\;\;\;\mbox{For any fixed }\alpha, M_{(\alpha +n)}=0\;\;\;\mbox{for
$n$ sufficiently small;}$

$(M8)\;\;\;\;\dim M_{(\alpha )}<\infty\;\;\;\mbox{for any }\alpha\in
{\bf C}.$

If $V$ is a vertex operator superalgebra, it is clear that any
$V$-module is a direct sum of $\frac{1}{2}{\bf Z}$-graded modules
which are truncated from below.

{\bf Remark 2.3.2}. As having been noticed by many authors (see for
example [HL]), the Virasoro algebra relation (V6) follows from the axioms
(M2)-(M4). This simply follows from Remark 2.2.3 and the commutator
formula (2.2.6).

It has been proved [FHL] that the
rationality, the commutativity and the associativity in terms of
``matrix coefficients'' are equivalent to the Jacobi
identity for vertex operator algebra and module. In fact, the part
three of the proof of Proposition 2.2.4 gives a proof without
involving  ``matrix-coefficients.'' Moreover, Proposition 2.2.6
proves the following proposition:

{\bf Proposition 2.3.3}. {\it Let $V$ be a
vertex superalgebra. Then the Jacobi identity for $V$-module can
be equivalently substituted by the associativity (2.2.9).}

{\bf Remark 2.3.4}. Since the
iterate formula (2.2.8) implies the associativity formula (2.2.9), it
follows from Proposition 2.3.3 that
the iterate formula (2.2.8) imply the Jacobi identity for module of a
vertex superalgebra.

For the rest of this subsection, we shall discuss some simple but
important analogues from the classical associative algebra theory,
which are useful in the study of module theory for some  vertex
operator algebras in Sections 4 and 5.

Let $V$ be a vertex superalgebra and let $M$
be a $V$-module. Let $a,b, c^{0},\cdots,c^{k}$ be $({\bf Z}/2{\bf
Z})$-homogeneous elements of $V$. If $M$ is faithful, it follows from
the commutator formula (2.2.6) and Lemma 2.1.4 that
\begin{eqnarray}
[Y_{M}(a,z_{1}),Y_{M}(b,z_{2})]=\sum_{i=0}^{k}\frac{1}{i!}z_{1}^{-i-1}
\delta^{(i)}\left(\frac{z_{2}}{z_{1}}\right)Y_{M}(c^{i},z_{2})
\end{eqnarray}
if and only if
\begin{eqnarray}
a_{i}b=0\;\;\mbox{for }i>k\;\mbox{ and }a_{i}b=c^{i}\;\;\mbox{for
}i=0,\cdots,k.
\end{eqnarray}
Then the following lemma is clear.

{\bf Lemma 2.3.5.} {\it Let $V$ be a vertex superalgebra and let $a,b,
c^{0},\cdots,c^{k}\in V$ be $({\bf Z}/2{\bf Z})$-homogeneous elements
such that}
\begin{eqnarray}
[Y_{V}(a,z_{1}),Y_{V}(b,z_{2})]=\sum_{i=0}^{k}\frac{1}{i!}z_{1}^{-i-1}
\delta^{(i)}\left(\frac{z_{2}}{z_{1}}\right)Y_{V}(c^{i},z_{2}).
\end{eqnarray}
{\it Then (2.3.2) holds for any $V$-module $M$.
Conversely, let $M$ be a faithful $V$-module such that (2.3.2) holds.
Then (2.3.4) holds.}

Noticing that $V$ is a faithful $V$-module, we  have:
$[Y(a,z_{1}),Y(b,z_{2})]_{\pm}=0$ if and only if $a_{i}b=0$ for all
$i\in {\bf Z}_{+}$.

{\bf Proposition 2.3.6 [DL].} {\it Let $V$ be any vertex operator
algebra and let $(W,Y_{W})$ be any $V$-module. Let
$v\in V$ be such that the component operators $v_{n}$ $(n\in {\bf Z})$
all commute with one another, so that $Y_{W}(v,z)^{N}$ is well defined
on $W$ for $N\in {\bf N}$. Then }
\begin{eqnarray}
Y_{W}((v_{-1})^{N}{\bf 1},z)=Y_{W}(v,z)^{N}.
\end{eqnarray}
{\it In particular, if $(v_{-1})^{N}{\bf 1}=0$ for a fixed $N$, then}
\begin{eqnarray}
Y_{W}(v,z)^{N}=0.\;\;\;\;\Box
\end{eqnarray}

{\bf Remark 2.3.7}. More generally, let $V$ be a vertex
superalgebra and let $v$ be an even or odd element of $V$ such that
$[Y_{W}(v,z_{1}),Y_{W}(v,z_{2})]_{\pm}=0$. Then the assertion of
Proposition 2.3.6 still holds. Conversely, let
$M$ be a faithful $V$-module and let $v\in V$ such that
$Y_{M}(v,z)^{N}=0$ on $M$ for $N\in {\bf N}$. Then $Y(v,z)^{N}=0$ on
$V$. The ``two-wedged'' formula (2.3.5)
turns out to be very useful in Section 5.

{\bf Remark 2.3.8.} Let $A$ be an associative algebra and let
$P(x_{1},\cdots,x_{n})$ be a polynomial in non-commuting variables $x_{i}$'s.
Let $a_{1},\cdots,a_{n}\in A$ be such that $P(a_{1},\cdots,a_{n})=0$
on $A$. Then $P(a_{1},\cdots,a_{n})=0$ on any $A$-module $M$.
Conversely, let $M$ be a faithful
$A$-module and let $a_{1},\cdots,a_{n}\in A$ such that
$P(a_{1},\cdots,a_{n})=0$ on $M$. Then $P(a_{1},\cdots,a_{n})=0$ on $A$.
{}From this point of view we may think of Lemma 2.3.5 and DL's
Proposition 2.3.6 as complex analogues for vertex (operator)
superalgebras of these classical facts.

\newpage

\section{Local systems of vertex operators}
In this section, we shall introduce what we call ``local systems of
vertex operators'' for any $({\bf Z}/2{\bf Z})$-graded vector space $M$
and present the main result of this paper. That is, any
local system has a natural vertex superalgebra structure
with $M$ as a module and for a fixed vertex
superalgebra $V$, giving a $V$-module $M$ is equivalent to giving a
vertex superalgebra homomorphism from $V$ to some local system of
vertex operators on $M$.

\subsection{Weak vertex operators}
Let us first define three categories $C,C^{o}$ and $C_{\ell}$. Define $C$
to be the category with $({\bf Z}/2{\bf Z})$-graded vector spaces as
its objects and with $({\bf Z}/2{\bf Z})$-homomorphisms as morphisms. We
define $C^{o}$ to be the category with $ob C^{o}$ consisting of
$(M,d)$ where $M$ is an object of $C$ and $d$ is a morphism from $M$
to $M$. A morphism from $(M,d_{1})$ to $(W,d_{2})$ is a morphism $f$
from $M$ to $W$ such that $fd_{1}=d_{2}f$. Recall that a
$Vir$-module $M$ is a {\it restricted} module if for any $u\in M$,
$L(n)u=0$ for $n$ sufficiently large.
For any complex
number $\ell$, we define $C_{\ell}$ to the category with restricted
$Vir$-modules of central charge $\ell$ as its objects and with
$Vir$-homomorphisms as morphisms.

Let $M=M^{0}\oplus M^{1}$ be a $({\bf Z}/2{\bf Z})$-graded vector
space. Then ${\rm End}M=({\rm End}M)^{0}\oplus ({\rm End}M)^{1}$ is
also a $({\bf Z}/2{\bf Z})$-graded vector space where
\begin{eqnarray}
({\rm End}M)^{0}&=&\{ A\in {\rm End}M| AM^{i}\subseteq M^{i}\;\mbox {
for }i=0,1\},\\
({\rm End}M)^{1}&=&\{ A\in {\rm End}M| AM^{0}\subseteq M^{1},
AM^{1}\subseteq M^{0}\}.
\end{eqnarray}
Furthermore,
\begin{eqnarray}
({\rm End}M)[[z,z^{-1}]]=({\rm
End}M)^{0}[[z,z^{-1}]]\oplus ({\rm End}M)^{1}[[z,z^{-1}]]
\end{eqnarray}
 is also a
$({\bf Z}/2{\bf Z})$-graded vector space. It is clear that the
derivative operator $\displaystyle{{d\over
dz}}$ is an endomorphism of $({\rm End}M)[[z,z^{-1}]]$, which preserves
both the even subspace and the odd subspace.

{\bf Definition 3.1.1.} Let $M$ be any $({\bf Z}/2{\bf Z})$-graded
vector space.  A {\it weak vertex operator} on $M$ is a
formal series $a(z)=\sum_{n \in {\bf
Z}}a_{n}z^{-n-1} \in ({\rm End}\:M)[[z,z^{-1}]]$ such that
\begin{eqnarray}
a(z)u\in M((z))\;\;\;\mbox{for any }u\in M.
\end{eqnarray}
That is, $a_{n}u=0$ for $n$ sufficiently large.
Let $(M,d)$ be a pair consisting of a $({\bf Z}/2{\bf Z})$-graded
vector space $M$ and a $({\bf Z}/2{\bf Z})$-endomorphism $d$ of $M$.
A {\it weak vertex operator
on $(M,d)$} is a weak vertex operator $a(z)$ on $M$ such that
\begin{eqnarray}
[d,a(z)]=a'(z) \left(={d\over dz}a(z)\right).
\end{eqnarray}

Denote by $F(M)$ (resp. $F(M,d)$) the space of all
weak vertex operators on $M$ (resp. $(M,d)$). It is clear that
both $F(M)$ and $F(M,d)$ are graded
subspaces of $({\rm End}M)[[z,z^{-1}]]$. It is
easy to see that $F(M)$ coincides with
${\rm Hom}_{{\bf C}}(M,M((z)))$.

By definition, it is clear that if $a(z)$ is a weak
vertex operator on
$M$ (resp. $(M,d)$), the formal derivative $a'(z)$ is also a weak
vertex operator on $M$ (resp. $(M,d)$). Then we have an endomorphism
$\displaystyle{D={d\over dz}}$ for both $F(M)$ and $F(M,d)$.

{\bf Definition 3.1.2.} Let $M$ be a restricted $Vir$-module of central
charge $\ell$. A
weak vertex operator $a(z)$ on $(M,L(-1))$ is
said to be of  {\it  weight} $h\in {\bf C}$  if it
satisfies the following condition:
\begin{eqnarray}
[L(0),a(z)]=ha(z)+za'(z).
\end{eqnarray}
Denote by $F(M,L(-1))_{(h)}$ the space of weak vertex operators
on $(M, L(-1))$ of weight $h$ and set
\begin{eqnarray}
F^{o}(M,L(-1))=\oplus_{h\in {\bf C}}F(M,L(-1))_{(h)}.
\end{eqnarray}

{\bf Remark 3.1.3}. For any super vector space $M$, the identity operator
$I(z)={\rm id}_{M}$ is a weak vertex operator on $M$.
Let $M$ be a restricted $Vir$-module. Then $I(z)={\rm
id}_{M}$ is a weak vertex operator on $(M, L(-1))$ of weight zero and
 $L(z)=\sum_ {n \in {\bf Z}}L(n)z^{-n-2}$
is a weak vertex operator on $(M,L(-1))$ of weight two.
If $a(z)$ is an even (resp. odd) weak vertex
operator on $(M,L(-1))$ of weight $h$, then
$\displaystyle{a'(z)={d\over dz}a(z)}$ is
an even (resp. odd) weak vertex operator of weight $h +1$.

{\bf Lemma 3.1.4}. {\it Let $M$ be a super vector space and let $a(z)$
and $b(z)$ be  $({\rm Z}/2{\bf Z})$-homogeneous weak vertex operators
on $M$. For any integer $n$, set}
\begin{eqnarray}
a(z)_{n}b(z)={\rm Res}_{z_{1}}\left((z_{1}-z)^{n}a(z_{1})b(z)
-(-1)^{|a(z)||b(z)|}(-z+z_{1})^{n}b(z)a(z_{1})\right).
\end{eqnarray}
{\it Then $a(z)_{n}b(z)$ is a $({\bf Z}/2{\bf Z})$-homogeneous weak
vertex operator satisfying $|a(z)_{n}b(z)|=|a(z)||b(z)|$.}

{\sl Proof.} For any $u\in M$, by definition we have
\begin{eqnarray}
& &(a(z)_{n}b(z))u\nonumber\\
&=&{\rm Res}_{z_{1}}\left((z_{1}-z)^{n}a(z_{1})b(z)u
-(-1)^{|a(z)||b(z)|}(-z+z_{1})^{n}b(z)a(z_{1})u\right)\nonumber\\
&=&\sum_{k=0}^{\infty}\left(\begin{array}{c}n\\k\end{array}\right)\left(
(-z)^{k}a_{n-k}b(z)u-(-1)^{|a(z)||b(z)|}(-z)^{n-k}b(z)a_{k}u\right).
\end{eqnarray}
It is easy to see that $(a(z)_{n}b(z))u\in M((z))$. Therefore,
$a(z)_{n}b(z)$ is a weak vertex operator on $M$.$\;\;\;\;\Box$

{\bf Definition 3.1.5}. Let $M$ be a super vector space and let
$a(z)$ and $b(z)$ be $({\rm Z}/2{\bf Z})$-homogeneous
weak vertex operators on $M$. Then we define
\begin{eqnarray}
& &Y(a(z),z_{0})b(z)\nonumber\\
&=&: \sum_{n\in {\bf Z}}a(z)_{n}b(z) z_{0}^{-n-1}\nonumber\\
&=&{\rm Res}_{z_{1}}
\left(z_{0}^{-1}\delta\left(\frac{z_{1}-z}{z_{0}}\right)a(z_{1})
b(z)-\varepsilon_{a,b}z_{0}^{-1}
\delta\left(\frac{z-z_{1}}{-z_{0}}\right)b(z)a(z_{1})\right).
\end{eqnarray}
Extending the definition bilinearly, we obtain a linear map
\begin{eqnarray}
Y(\cdot,z_{0}) :& & F(M)\rightarrow ({\rm End}
F(M))[[z_{0},z_{0}^{-1}]];\nonumber\\
 & &a(z)\mapsto Y(a(z),z_{0}).
\end{eqnarray}

{\bf Lemma 3.1.6}. {\it For any $a(z)\in F(M)$, we have}
\begin{eqnarray}
& &Y(I(z),z_{0})a(z)=a(z);\\
& &Y(a(z),z_{0})I(z)=e^{z_{0}{\partial\over\partial
z}}a(z)\;(=a(z+z_{0})).
\end{eqnarray}

{\sl Proof}. By definition, we have:
\begin{eqnarray*}
& &Y(I(z),z_{0})a(z)\\
&=&{\rm Res}_{z_{1}}\left(z_{0}^{-1}\delta\left(\frac{z_{1}-z}{z_{0}}
\right)a(z)-z_{0}^{-1}\delta\left(\frac{-z+z_{1}}{z_{0}}\right)
a(z)\right)\\
&=&{\rm Res}_{z_{1}}z^{-1}\delta\left(\frac{z_{1}-z_{0}}{z}\right)
a(z)\\
&=&a(z)
\end{eqnarray*}
and
\begin{eqnarray}
& &Y(a(z),z_{0})I(z)\nonumber\\
&=&{\rm Res}_{z_{1}}
\left(z_{0}^{-1}\delta\left(\frac{z_{1}-z}{z_{0}}\right)a(z_{1})I(z)-
z_{0}^{-1}\delta\left(\frac{-z+z_{1}}{z_{0}}\right)I(z)a(z_{1})
\right)\nonumber\\
&=&{\rm Res}_{z_{1}}
z^{-1}\delta\left(\frac{z_{1}-z_{0}}{z}\right)a(z_{1})\nonumber\\
&=&{\rm Res}_{z_{1}}z_{1}^{-1}\delta\left(\frac{z+z_{0}}
{z_{1}}\right)a(z_{1})\nonumber\\
&=&{\rm Res}_{z_{1}}z_{1}^{-1}\delta\left(\frac{z+z_{0}}
{z_{1}}\right)a(z+z_{0})\nonumber\\
&=&a(z+z_{0})\nonumber\\
&=&e^{z_{0}{\partial\over\partial z}}a(z).\;\;\;\;\Box
\end{eqnarray}

{\bf Lemma 3.1.7}. {\it Let $M\in ob C$ and  $a(z),b(z)\in F(M)$.
Then we have}
\begin{eqnarray}
{\partial\over \partial z_{0}}Y(a(z),z_{0})b(z)=
Y(D(a(z)),z_{0})b(z)=[D,Y(a(z),z_{0})]b(z).
\end{eqnarray}

{\sl Proof.} Without losing generality, we may assume that both $a(z)$
and $b(z)$ are $({\bf Z}/2{\bf Z})$-homogeneous. By definition, we have
\begin{eqnarray}
& &{\partial\over \partial z_{0}}Y(a(z),z_{0})b(z)\nonumber\\
&=&{\rm Res}_{z_{1}}\left({\partial \over \partial z_{0}}\left( z_{0}^{-1}
\delta\left(\frac{z_{1}-z}{z_{0}}\right)\right)
a(z_{1})b(z)-\varepsilon_{a,b}{\partial \over \partial z_{0}}\left(
z_{0}^{-1}\delta\left(\frac{-z+z_{1}}{z_{0}}\right)\right)b(z)a(z_{1})
\right)\nonumber\\
&=&-{\rm Res}_{z_{1}}
\left({\partial\over \partial z_{1}} z_{0}^{-1}\delta\left(
\frac{z_{1}-z}{z_{0}}\right)\right)a(z_{1})b(z)\nonumber\\
& &+{\rm Res}_{z_{1}}\varepsilon_{a,b}\left({\partial\over \partial z_{1}}
z_{0}^{-1}
\delta\left(\frac{-z+z_{1}}{z_{0}}\right)\right)b(z)a(z_{1})\;\;
\mbox{(by Lemma 2.1.1)}\nonumber\\
&=&{\rm Res}_{z_{1}}
\left( z_{0}^{-1}\delta\left(\frac{z_{1}-z}{z_{0}}\right)a'(z_{1})
b(z)-\varepsilon_{a,b}z_{0}^{-1}
\delta\left(\frac{-z+z_{1}}{z_{0}}\right)b(z)a'(z_{1})\right)
\nonumber\\
&=&Y(a'(z),z_{0})b(z)
\end{eqnarray}
and
\begin{eqnarray}
& &[D,Y(a(z),z_{0})]b(z)\nonumber \\
&=&D(Y(a(z),z_{0})b(z))-Y(a(z),z_{0})Db(z)\nonumber \\
&=&{\partial \over \partial
z}(Y(a(z),z_{0})b(z))-Y(a(z),z_{0})b'(z)\nonumber \\
&=&{\rm Res}_{z_{1}}\left(\left({\partial \over \partial z}z_{0}^{-1}
\delta\left(\frac{z_{1}-z}{z_{0}}\right)\right)a(z_{1})b(z)
-\varepsilon_{a,b}\left({\partial \over \partial z}z_{0}^{-1}
\delta\left(\frac{z-z_{1}}{-z_{0}}\right)\right)b(z)a(z_{1})\right)
\nonumber\\
&=&{\rm Res}_{z_{1}}\left({\partial \over \partial z_{0}}\left(z_{0}^{-1}
\delta\left(\frac{z_{1}-z}{z_{0}}\right)\right)a(z_{1})b(z)
-\varepsilon_{a,b}{\partial \over \partial z_{0}}\left(z_{0}^{-1}
\delta\left(\frac{z-z_{1}}{-z_{0}}\right)\right)b(z)a(z_{1})\right)
\nonumber\\
&=&{\partial \over \partial z_{0}}Y(a(z),z_{0})b(z)\nonumber\\
&=&Y(a'(z),z_{0})b(z)\nonumber\\
&=&Y(D\cdot a(z),z_{0})b(z).\;\;\;\;\Box
\end{eqnarray}

{\bf Lemma 3.1.8}.  {\it Let $(M,d)$ be an object of $C^{o}$ and
let $a(z),b(z)\in F(M,d)$. Then $a(z)_{n} b(z)\in F(M,d)$.
Furthermore, if $M$ is a restricted Vir-module with
central charge $\ell$ and $a(z),b(z)$ are weak vertex operators on
$(M,L(-1))$ of weights $\alpha, \beta$, respectively, then for any
integer $n$, $a(z)_{n} b(z)$ is a weak vertex
operator of weight $(\alpha +\beta -n-1)$ on $(M,L(-1))$}.

{\sl Proof.} It is equivalent to prove the following:
\begin{eqnarray}
& &[L(-1),Y(a(z),z_{0})b(z)]={\partial \over \partial
z}(Y(a(z),z_{0})b(z));\\
& &[L(0),Y(a(z),z_{0})b(z)]\nonumber\\
&=&(\alpha +\beta)Y(a(z),z_{0})b(z)+z_{0}{\partial\over \partial
z_{0}}(Y(a(z),z_{0})b(z))+z{\partial\over \partial
z}(Y(a(z),z_{0})b(z)).
\end{eqnarray}
Without losing generality we may assume that both $a(z)$ and $b(z)$ are
homogeneous. By definition, we have:
\begin{eqnarray}
& &{\partial\over \partial z}(Y(a(z),z_{0})b(z))\nonumber\\
&=&{\partial\over \partial z}{\rm Res}_{z_{1}}
\left(z_{0}^{-1}\delta\left(\frac{z_{1}-z}{z_{0}}\right)a(z_{1})b(z)
-\varepsilon_{a,b}z_{0}^{-1}\delta\left(\frac{z-z_{1}}{-z_{0}}\right)
b(z)a(z_{1})\right)\nonumber\\
&=&{\rm Res}_{z_{1}}\left(\left({\partial\over\partial z}z_{0}^{-1}
\delta \left(\frac{z_{1}-z}{z_{0}}\right)\right)a(z_{1})b(z)
-\varepsilon_{a,b}\left({\partial\over\partial z}z_{0}^{-1}
\delta \left(\frac{z-z_{1}}{-z_{0}}\right)\right)
b(z)a(z_{1})\right)\nonumber\\
& &+{\rm Res}_{z_{1}}
\left(\delta\left(\frac{z_{1}-z}{z_{0}}\right)a(z_{1})b'(z)
-\varepsilon_{a,b}z_{0}^{-1}\delta\left(\frac{z-z_{1}}{-z_{0}}\right)
b'(z)a(z_{1})\right)\nonumber\\
&=&-{\rm Res}_{z_{1}}
\left(\left({\partial\over \partial z_{1}}z_{0}^{-1}\delta\left(\frac
{z_{1}-z}{z_{0}}\right)\right)a(z_{1})b(z)-\varepsilon_{a,b}
\left({\partial\over \partial z_{1}}z_{0}^{-1}\delta\left(\frac{z_{1}-z}
{z_{0}}\right)\right)b(z)a(z_{1})\right)\nonumber\\
& &+{\rm Res}_{z_{1}}\left(z_{0}^{-1}\delta\left(\frac{z_{1}-z}
{z_{0}}\right)a(z_{1})b'(z)
-\varepsilon_{a,b}z_{0}^{-1}\delta \left(\frac{z_{1}-z}{z_{0}}\right)
b'(z)a(z_{1})\right)\nonumber\\
&=&{\rm Res}_{z_{1}}
\left(z_{0}^{-1}\delta\left(\frac{z_{1}-z}{z_{0}}\right)a(z_{1})b'(z)
-\varepsilon_{a,b}z_{0}^{-1}\delta \left(\frac{z-z_{1}}{-z_{0}}\right)
b'(z)a(z_{1})\right)\nonumber\\
& &-{\rm Res}_{z_{1}}
\left(z_{0}^{-1}\delta\left(\frac{z_{1}-z}{z_{0}}\right)a'(z_{1})b(z)
-\varepsilon_{a,b}z_{0}^{-1}\delta \left(\frac{z-z_{1}}{-z_{0}}\right)
b(z)a'(z_{1})\right)\nonumber\\
&=&[L(-1),Y(a(z),z_{0})b(z)] \nonumber\end{eqnarray}
and
\begin{eqnarray}
& &[L(0),Y(a(z),z_{0})b(z)]\nonumber\\
&=&{\rm Res}_{z_{1}}
\left(z_{0}^{-1}\delta\left(\frac{z_{1}-z}
{z_{0}}\right)[L(0),a(z_{1})b(z)]
-\varepsilon_{a,b}z_{0}^{-1}\delta \left(\frac{z-z_{1}}{-z_{0}}\right)
[L(0),b(z)a(z_{1})]\right)\nonumber\\
&=&{\rm Res}_{z_{1}}z_{0}^{-1}\delta\left(\frac{z_{1}-z}
{z_{0}}\right)\left(a(z_{1})[L(0),b(z)]+[L(0),a(z_{1})]b(z)\right)
\nonumber\\
& & -{\rm Res}_{z_{1}}\varepsilon_{a,b}z_{0}^{-1}\delta\left(
\frac{z-z_{1}}{-z_{0}}\right)
\left(b(z)[L(0),a(z_{1})]+[L(0),b(z)]a(z_{1})\right)\nonumber\\
&=&{\rm Res}_{z_{1}}z_{0}^{-1}\delta\left(\frac{z_{1}-z}
{z_{0}}\right)\left(\beta a(z_{1})b(z)+za(z_{1})b'(z)+\alpha
a(z_{1})b(z)+z_{1}a'(z_{1})b(z)\right)\nonumber\\
& &-{\rm Res}_{z_{1}}\varepsilon_{a,b}z_{0}^{-1}\delta\left(\frac{z-z_{1}}
{-z_{0}}\right)\left(\alpha b(z)a(z_{1})+z_{1}b(z)a'(z_{1})
+\beta b(z)a(z_{1})+zb'(z)a(z_{1})\right)\nonumber\\
&=&(\alpha +\beta )Y(a(z),z_{0})b(z)\nonumber\\
& &+{\rm Res}_{z_{1}}\left(z_{0}^{-1}\delta\left(\frac{z_{1}-z}
{z_{0}}\right)za(z_{1})b'(z)-\varepsilon_{a,b}z_{0}^{-1}\delta\left(
\frac{z-z_{1}}{-z_{0}}\right)zb'(z)a(z_{1})\right)\nonumber\\
& &-{\rm Res}_{z_{1}}\left(\left({\partial\over\partial z_{1}}z_{1}z_{0}^{-1}
\delta\left(\frac{z_{1}-z}{z_{0}}\right)\right)a(z_{1})b(z)
-\varepsilon_{a,b}\left({\partial\over\partial z_{1}}z_{1}z_{0}^{-1}
\delta\left(\frac{z-z_{1}}{-z_{0}}\right)\right)b(z)a(z_{1})\right)
\nonumber\\
&=&(\alpha +\beta )Y(a(z),z_{0})b(z)\nonumber\\
& &+{\rm Res}_{z_{1}}\left(z_{0}^{-1}\delta\left(\frac{z_{1}-z}
{z_{0}}\right)za(z_{1})b'(z)-\varepsilon_{a,b}z_{0}^{-1}\delta\left(
\frac{z-z_{1}}{-z_{0}}\right)zb'(z)a(z_{1})\right)\nonumber\\
& &-{\rm Res}_{z_{1}}(z_{0}+z)\left({\partial\over\partial z_{1}}
z_{0}^{-1}\delta\left(\frac{z_{1}-z}{z_{0}}\right)\right)a(z_{1})b(z)
\nonumber\\
& &+ {\rm Res}_{z_{1}}(z_{0}+z)\varepsilon_{a,b}\left({\partial\over
\partial z_{1}}z_{0}^{-1}
\delta\left(\frac{z-z_{1}}{-z_{0}}\right)\right)b(z)a(z_{1})
\nonumber\\
&=&(\alpha +\beta )Y(a(z),z_{0})b(z)\nonumber\\
& &+{\rm Res}_{z_{1}}\left(z_{0}^{-1}\delta\left(\frac{z_{1}-z}
{z_{0}}\right)za(z_{1})b'(z)-\varepsilon_{a,b}z_{0}^{-1}\delta\left(
\frac{z-z_{1}}{-z_{0}}\right)zb'(z)a(z_{1})\right)\nonumber\\
& &+{\rm Res}_{z_{1}}z_{0}\left(\left({\partial\over\partial z_{0}}
z_{0}^{-1}\delta\left(\frac{z_{1}-z}{z_{0}}\right)\right)a(z_{1})b(z)
-\varepsilon_{a,b}\left({\partial\over\partial z_{1}}z_{0}^{-1}
\delta\left(\frac{z-z_{1}}{-z_{0}}\right)\right)b(z)a(z_{1})\right)
\nonumber\\
& &+{\rm Res}_{z_{1}}z\left(\left({\partial\over\partial z}
z_{0}^{-1}\delta\left(\frac{z_{1}-z}{z_{0}}\right)\right)a(z_{1})b(z)
-\varepsilon_{a,b}\left({\partial\over\partial z_{1}}z_{0}^{-1}
\delta\left(\frac{z-z_{1}}{-z_{0}}\right)\right)b(z)a(z_{1})\right)
\nonumber\\
&=&(\alpha +\beta )Y(a(z),z_{0})b(z)+z{\partial\over \partial z}
\left(Y(a(z),z_{0})b(z)\right)+z_{0}{\partial\over \partial z_{0}}\left(
Y(a(z),z_{0})b(z)\right).\nonumber\;\;\;\;\Box
\end{eqnarray}

\subsection{Vertex operators and local systems}
The following definition is motivated by physicists' work for example [Go].

{\bf Definition 3.2.1}. Two $({\bf Z}/2{\bf Z})$-homogeneous weak
vertex operators $a(z_{1})$ and $b(z_{2})$ are
said to be {\it mutually local} if there a positive integer $n$ such that
\begin{eqnarray}
(z_{1}-z_{2})^{n}a(z_{1})b(z_{2})=(-1)^{|a(z)|b(z)|}(z_{1}-z_{2})^{n}
b(z_{2})a(z_{1}).
\end{eqnarray}
A $({\bf Z}/2{\bf Z})$-homogeneous weak vertex operator is called a
{\it vertex operator} if it is
local with itself, a graded subspace $A$ of $F(M)$ is
said to be {\it local} if any two $({\bf Z}/2{\bf Z})$-homogeneous
weak vertex operators in $A$ are
mutually local, and {\it a local system} of vertex operators on $M$ is
a maximal local (graded) space of $F(M)$.

{\bf Remark 3.2.2}. Let $V$ be a vertex superalgebra and let $(M, Y_{M})$ be a
$V$-module. Then the image of $V$ under the linear map
$Y_{M}(\cdot,z)$ is a local subspace of $F(M)$.

{\bf Remark 3.2.3}. Let $M$ be a super vector space and let $a(z)$ and
 $b(z)$ be homogeneous mutually local weak vertex operators on $M$.
Let $k$ be a positive integer satisfying (3.2.1). Then
$a(z)_{n}b(z)=0$ whenever $n\ge k$. Thus $Y(a(z),z_{0})b(z)$
involves only finitely many negative powers of $z_{0}$. (This
corresponds to the truncation condition (V1).)

{\bf Lemma 3.2.4}. {\it If $a(z_{1})$ is local with $b(z_{2})$, then
$a(z_{1})$ is local with $b'(z_{2})$}.

{\sl Proof}. Let $n$ be a positive integer such that (3.2.1) holds. Then
\begin{eqnarray}
(z_{1}-z_{2})^{n+1}a(z_{1})b(z_{2})=(-1)^{|a(z)|b(z)|}(z_{1}-z_{2})^{n+1}
b(z_{2})a(z_{1}).
\end{eqnarray}
Differentiating (3.2.2) with respect to $z_{2}$, then using (3.2.1) we obtain
\begin{eqnarray}
(z_{1}-z_{2})^{n+1}a(z_{1})b'(z_{2})=(-1)^{|a(z)|b(z)|}(z_{1}-z_{2})^{n+1}
b'(z_{2})a(z_{1}).\;\;\;\;\Box
\end{eqnarray}

{\bf Remark 3.2.5}. For any super vector space $M$, it follows from
Zorn's lemma that there always
exist local systems of vertex operators on $M$. Since the identity
operator $I(z)={\rm id}_{M}$ is mutually local with any weak vertex
operator on $M$, any local system contains $I(z)$.
{}From Remark 3.1.3 and Lemma 3.2.4, any local system
is closed under the derivative operator $\displaystyle{D={d\over dz}}$.

{\bf Lemma 3.2.6}. {\it Let $M$ be a restricted $Vir$-module with
central charge
$\ell$. Then $L(z)=\sum_{n\in {\bf Z}}L(n)z^{-n-2}$ is an even (local)
vertex operator on $(M,L(-1))$ of weight two.}

{\sl Proof}. It follows from Remark 2.2.3 and
 Lemma 2.1.3 that
\begin{eqnarray}
(z_{1}-z_{2})^{k}[L(z_{1}),L(z_{2})]=0\;\;\;\mbox{ for }k\ge 4.
\end{eqnarray}
Then $L(z)$ is a local vertex operator on $M$. $\;\;\;\;\Box$

The proof of the following proposition was given by Professor Chongying Dong.

{\bf Proposition 3.2.7}. {\it Let $a(z)$, $b(z)$ and $c(z)$ be
$({\bf Z}/2{\bf Z})$-homogeneous weak vertex operators on $M$. Suppose
both $a(z)$ and $b(z)$ are local with $c(z)$. Then
$a(z)_{n}b(z)$ is local with $c(z)$ for all $n\in {\bf Z}$}.

{\sl Proof}. Let $r$ be a positive integer greater than $-n$ such that the
following identities hold:
\begin{eqnarray*}
&
&(z_{1}-z_{2})^{r}a(z_{1})b(z_{2})=\varepsilon_{a,b}(z_{1}-z_{2})^{r}b(z_{2})a(z_{1}),\\
&
&(z_{1}-z_{2})^{r}a(z_{1})c(z_{2})=\varepsilon_{a,c}(z_{1}-z_{2})^{r}c(z_{2})a(z_{1}),\\
&
&(z_{1}-z_{2})^{r}b(z_{1})c(z_{2})=\varepsilon_{b,c}(z_{1}-z_{2})^{r}c(z_{2})b(z_{1}).
\end{eqnarray*}
By definition, we have
\begin{eqnarray}
a(z)_{n}b(z)
={\rm Res}_{z_{1}}\left((z_{1}-z)^{n}a(z_{1})b(z)
-\varepsilon_{a,b}(-z+z_{1})^{n}b(z)a(z_{1})\right).
\end{eqnarray}
Since
\begin{eqnarray}
& &(z-z_{3})^{4r}\left((z_{1}-z)^{n}a(z_{1})b(z)c(z_{3})
-\varepsilon_{a,b}(-z+z_{1})^{n}b(z)a(z_{1})c(z_{3})\right)\nonumber\\
&=&\sum_{s=0}^{3r}\left(\begin{array}{c}3r\\s\end{array}\right)
(z-z_{1})^{3r-s}(z_{1}-z_{3})^{s}(z-z_{3})^{r}\cdot
\nonumber\\
& &\cdot \left((z_{1}-z)^{n}a(z_{1})b(z)c(z_{3})
-\varepsilon_{a,b}(-z+z_{1})^{n}b(z)a(z_{1})c(z_{3})\right)\nonumber\\
&=&\sum_{s=r+1}^{3r}\left(\begin{array}{c}3r\\s\end{array}\right)
(z-z_{1})^{3r-s}(z_{1}-z_{3})^{s}(z-z_{3})^{r}\cdot
\nonumber\\
& &\cdot \left((z_{1}-z)^{n}a(z_{1})b(z)c(z_{3})
-\varepsilon_{a,b}(-z+z_{1})^{n}b(z)a(z_{1})c(z_{3})\right)\nonumber\\
&=&\sum_{s=r+1}^{3r}\left(\begin{array}{c}3r\\s\end{array}\right)
(z-z_{1})^{3r-s}(z_{1}-z_{3})^{s}(z-z_{3})^{r}\cdot
\nonumber\\
& &\cdot \varepsilon_{a,c}\varepsilon_{b,c} \left((z_{1}-z)^{n}c(z_{3})a(z_{1})
b(z)
-\varepsilon_{a,b}(-z+z_{1})^{n}c(z_{3})b(z)a(z_{1})\right)\nonumber\\
&=&\varepsilon_{a,c}\varepsilon_{b,c}(z-z_{3})^{4r}\left((z_{1}-z)^{n}c(z_{3})a(z_{1})b(z)
-\varepsilon_{a,b}(-z+z_{1})^{n}c(z_{3})b(z)a(z_{1})\right),\nonumber\\
& &\mbox{}
\end{eqnarray}
we have
\begin{eqnarray}
(z-z_{3})^{4r}(a(z)_{n}b(z))c(z_{3})=\varepsilon_{a,c}\varepsilon_{b,c}
(z-z_{3})^{4r}c(z_{3})
(a(z)_{n}b(z)).\;\;\;\;\Box
\end{eqnarray}

{\bf Remark 3.2.8.} Let $M$ be any super vector space and let $V$ be
any local system of vertex operators on $M$. Then it follows from
Proposition 3.2.7, Remarks 3.2.3 and 3.2.5 and Lemmas 3.1.6 and 3.1.7
that the quadruple $(V,I(z),D, Y)$ satisfies (V1)-(V4) of Definition 2.2.1.

{\bf Proposition 3.2.9}. {\it Let $V$ be any local system of
vertex operators on $M$. Then
for any  vertex operators $a(z)$ and $b(z)$ in $V$,
$Y(a(z),z_{1})$ and $Y(b(z),z_{2})$ are mutually local
on $(V,D)$.}

{\sl Proof}. Let $c(z)$ be any homogeneous weak vertex operator on
$M$. Then we have
\begin{eqnarray*}
& &Y(a(z),z_{3})Y(b(z),z_{0})c(z_{2})\\
&=&{\rm Res}_{z_{1}}
z_{3}^{-1}\delta\left(\frac{z_{1}-z_{2}}{z_{3}}\right)
a(z_{1})(Y(b(z),z_{0})c(z_{2}))\\
&
&-\varepsilon_{a,b}\varepsilon_{a,c}z_{3}^{-1}\delta\left(\frac{-z_{2}+z_{1}}{z_{3}}\right)
(Y(b(z),z_{0})c(z_{2}))a(z_{1})\\
&=&{\rm Res}_{z_{1}}{\rm Res}_{z_{4}}A
\end{eqnarray*}
where
\begin{eqnarray*}
A&=& z_{3}^{-1}\delta\left(\frac{z_{1}-z_{2}}{z_{3}}\right)z_{0}^{-1}
\delta\left(\frac{z_{4}-z_{2}}{z_{0}}\right)a(z_{1})b(z_{4})c(z_{2})\\
& &-\varepsilon_{b,c} z_{3}^{-1}\delta\left(\frac{z_{1}-z_{2}}{z_{3}}\right)
z_{0}^{-1}
\delta\left(\frac{-z_{2}+z_{4}}{z_{0}}\right)a(z_{1})c(z_{2})b(z_{4})\\
& &-\varepsilon_{a,b}\varepsilon_{a,c}z_{3}^{-1}\delta\left(
\frac{-z_{2}+z_{1}}{z_{3}}\right)z_{0}^{-1}
\delta\left(\frac{z_{4}-z_{2}}{z_{0}}\right)b(z_{4})c(z_{2})a(z_{1})\\
& &+\varepsilon_{a,b}\varepsilon_{a,c}\varepsilon_{b,c}
z_{3}^{-1}\delta\left(\frac{-z_{2}+z_{1}}{z_{3}}\right)z_{0}^{-1}
\delta\left(\frac{-z_{2}+z_{4}}{z_{0}}\right)c(z_{2})b(z_{4})a(z_{1}).
\end{eqnarray*}
Similarly, we have
\begin{eqnarray}
Y(b(z),z_{0})Y(a(z),z_{3})c(z_{2})
={\rm Res}_{z_{1}}{\rm Res}_{z_{4}}B
\end{eqnarray}
where
\begin{eqnarray*}
B&=& z_{3}^{-1}\delta\left(\frac{z_{1}-z_{2}}{z_{3}}\right)z_{0}^{-1}
\delta\left(\frac{z_{4}-z_{2}}{z_{0}}\right)b(z_{4})a(z_{1})c(z_{2})\\
& &-\varepsilon_{a,c}z_{3}^{-1}\delta\left(\frac{-z_{2}+z_{1}}{z_{3}}\right)
z_{0}^{-1}
\delta\left(\frac{z_{4}-z_{2}}{z_{0}}\right)b(z_{4})c(z_{2})a(z_{1})\\
& &-
\varepsilon_{a,b}\varepsilon_{b,c}z_{3}^{-1}\delta\left(\frac{z_{1}-z_{2}}{z_{3}}\right)z_{0}^{-1}
\delta\left(\frac{-z_{2}+z_{4}}{z_{0}}\right)a(z_{1})c(z_{2})b(z_{4})\\
&
&+\varepsilon_{a,b}\varepsilon_{b,c}\varepsilon_{a,c}z_{3}^{-1}\delta\left(\frac{-z_{2}+z_{1}}{z_{3}}\right)z_{0}^{-1}
\delta\left(\frac{-z_{2}+z_{4}}{z_{0}}\right)c(z_{2})a(z_{1})b(z_{4}).
\end{eqnarray*}
Let $k$ be any positive integer such that
\begin{eqnarray*}
(z_{1}-z_{4})^{k}a(z_{1})b(z_{4})=\varepsilon_{a,b}(z_{1}-z_{4})^{k}b(z_{4})a(z_{1}).
\end{eqnarray*}
Since
\begin{eqnarray*}
(z_{3}-z_{0})^{k}z_{3}^{-1}\delta\left(\frac{z_{1}-z_{2}}{z_{3}}\right)z_{0}^{-1}
\delta\left(\frac{z_{4}-z_{2}}{z_{0}}\right)
=(z_{1}-z_{4})^{k}z_{3}^{-1}\delta\left(\frac{z_{1}-z_{2}}{z_{3}}\right)z_{0}^{-1}
\delta\left(\frac{z_{4}-z_{2}}{z_{0}}\right),
\end{eqnarray*}
it is clear that
locality of $a(z)$ with $b(z)$ implies the locality of $Y(a(z),z_{1})$
with $Y(b(z),z_{2}).\;\;\;\;\Box$

Now, we are ready to present our main theorem:

{\bf Theorem 3.2.10}. {\it Let $M$ be any $({\bf Z}/2{\bf Z})$-graded
vector space and let $V$ be any local system of vertex
operators on $M$. Then $V$ is a vertex superalgebra and $M$ satisfies
all the conditions for module except the existence of $d$ in (M2). If
$V$ is a local system on $(M,d)$, then $(M,d)$ is  a $V$-module.}

{\sl Proof}. It follows from Proposition 2.2.4, Remark 3.2.8 and
Proposition 3.2.9 that
$V$ is a vertex superalgebra. It follows from Proposition 2.3.3 and
Remark 2.3.3 that $M$ is a
$V$-module through the linear map $Y_{M}(a(z),z_{0})=a(z_{0})$ for
$a(z)\in V. \;\;\;\;\Box$

{\bf Corollary 3.2.11.} {\it Let $M$ be any $({\bf Z}/2{\bf Z})$-graded vector
space and let $S$ be any set of mutually local homogeneous  vertex operators
on $M$. Let $<S>$ be the subspace of $F(M)$ generated by $S\cup \{I(z)\}$
under the vertex operator multiplication (3.1.10) (or (3.1.8) for
components). Then $(<S>, I(z), D,Y)$ is a vertex superalgebra with $M$
as a module.}

{\sl Proof.} It follows from Proposition 3.2.7 that
$<S>$ is a local subspace of $ F(M)$. Let $A$ be a local system
containing $<S>$ as a subspace. Then by Theorem 3.2.10, $A$ is a vertex
superalgebra with $M$ as a module. Since $<S>$ is closed under
(3.2.10), $<S>$ is a vertex subalgebra. Since the ``multiplication''
(3.1.10) does not depend on the choice of the local system $A$,
$\langle S\rangle$ is canonical.$\;\;\;\;\Box$

{\bf Proposition 3.2.12}. {\it Let $M$ be a restricted $Vir$-module with
central charge $\ell$ and let $V$ be a
local system of
vertex operators on $(M, L(-1))$, containing $L(z)$. Then
the vertex operator $L(z)$
is a Virasoro element of the vertex superalgebra $V$.}

{\sl Proof}. First, by Theorem 3.2.10 $V$ is a vertex superalgebra with
$M$ as  a
$V$-module. Set $\omega=L(z)\in V$. By Lemma 2.3.7, the components of
vertex operator $Y(\omega,z_{0})$  give rise to a representation
on $V$ of central charge $\ell$ for the Virasoro algebra $Vir$.
For any
$a(z)\in V_{(h)}$, by definition we have
\begin{eqnarray}
& &L(z)_{0}a(z)=[L(-1),a(z)]=a'(z);\\
& &L(z)_{1}a(z)=[L(0),a(z)]-z^{-1}[L(-1),a(z)]=h a(z).
\end{eqnarray}
Therefore $V$ satisfies all conditions for a vertex
operator superalgebra except the requirements on the homogeneous
subspaces.$\;\;\;\;\Box$

Let $V$ be a vertex (operator) superalgebra and let $(M,d)$ be a
$V$-module. Then the image $\bar{V}$ of $V$ inside $F(M,d)$ is a
graded local subspace. By Zorn's lemma, there exists a local system
$A$ containing $\bar{V}$ as a subspace. From the vacuum property (M2)
we have:
\begin{eqnarray}
Y_{M}(\cdot,z)({\bf 1})=Y_{M}({\bf 1},z)={\rm id}_{M}=I(z).
\end{eqnarray}
For any $({\bf Z}/2{\bf Z})$-homogeneous elements $a,b\in V$, we have:
\begin{eqnarray}
& &Y_{M}(\cdot,z)(Y(a,z_{0})b)\nonumber\\
&=&Y_{M}(Y(a,z_{0})b,z)\nonumber\\
&=&{\rm Res}_{z_{1}}\left(z_{0}^{-1}\delta\left(\frac{z_{1}-z}{z_{0}}\right)
Y_{M}(a,z_{1})
Y_{M}(b,z)-\varepsilon_{a,b}z_{0}^{-1}\delta\left(\frac{z-z_{1}}
{-z_{0}}\right)Y_{M}(b,z)Y_{M}(a,z_{1})\right)\nonumber\\
&=&Y_{A}(Y_{M}(a,z),z_{0})Y_{M}(b,z).
\end{eqnarray}
Thus $Y_{M}(\cdot,z)$ is a vertex superalgebra homomorphism from $V$
to $A$. Conversely, let $\phi$ be a vertex superalgebra homomorphism from $V$
to some local system $A$ of vertex operators on $(M,d)$. Since $\phi
(u)\in A\in ({\rm End}M)[[z,z^{-1}]]$ for any $u\in V$, we use
$\phi_{z}$ for $\phi$ to indicate the dependence of $\phi(u)$ on $z$.
For any formal variable $z_{1}$, set
$\phi_{z_{1}}(a)=\phi_{z}(a)|_{z=z_{1}}$ for any $a\in V$.
We define
$Y_{M}(a,z)u=\phi_{z} (a)$ for $a\in V$. By definition we have:
\begin{eqnarray}
Y_{M}({\bf 1},z)=\phi_{z}({\bf 1})=I(z)={\rm id}_{M}.
\end{eqnarray}
For $({\bf Z}/2{\bf Z})$-homogeneous elements $a,b\in V$, we have:
\begin{eqnarray}
& &Y_{M}(Y(a,z_{0})b,z_{2})\nonumber\\
&=&\phi_{z} (Y(a,z_{0})b)|_{z=z_{2}}\nonumber\\
&=&\left(Y_{A}(\phi_{z}(a),z_{0})\phi_{z}(b)\right)|_{z=z_{2}}\nonumber\\
&=&{\rm Res}_{z_{1}}\left(z_{0}^{-1}\delta\left(\frac{z_{1}-z_{2}}
{z_{0}}\right)\phi_{z_{1}}(a)
\phi_{z_{2}}(b)-\varepsilon_{a,b}z_{0}^{-1}\delta\left(\frac{z_{2}-z_{1}}
{-z_{0}}\right)\phi_{z_{2}}(b)\phi_{z_{1}}(a)\right)\nonumber\\
&=&{\rm Res}_{z_{1}}\left(z_{0}^{-1}\delta\left(\frac{z_{1}-z_{2}}{z_{0}}
\right)Y_{M}(a,z_{1})
Y_{M}(b,z_{2})-\varepsilon_{a,b}z_{0}^{-1}\delta\left(\frac{z_{2}-z_{1}}
{-z_{0}}\right)Y_{M}(b,z_{2})Y_{M}(a,z_{1})\right).\nonumber\\
& &\mbox{ }
\end{eqnarray}
It follows from Remark 2.3.4 that $(M,d,Y_{M})$ is a $V$-module.
Therefore, we have proved:

{\bf Proposition 3.2.13.} {\it Let $V$ be a vertex (operator)
superalgebra. Then
giving a $V$-module $(M,d)$ is equivalent to giving a vertex superalgebra
homomorphism from $V$ to some local system of vertex operators on $(M,d)$.}

\newpage
\section{Vertex operator superalgebras and modules
associated to some infinite-dimensional affine Lie superalgebras}
In this section, we shall use the machinery we built in Section 3 to
study vertex operator superalgebras and modules associated
to the representations for some well-known infinite-dimensional Lie algebras
 or Lie superalgebras such as the Virasoro algebra, the Neveu-Schwarz
algebra and affine Lie superalgebras.

\subsection{Vertex operator algebras associated to the Virasoro algebra}
Let us start with an abstract result which will be used in this
section and the next section.

{\bf Proposition 4.1.1}. {\it Let $(V,D, {\bf 1},Y)$ be a vertex
superalgebra and let $(M,d,Y_{M})$ be a $V$-module. Let
$u\in M$ such that $du=0$. Then the linear map}
\begin{eqnarray}
f: V\rightarrow M; a\mapsto a_{-1}u\;\;\;\mbox{ {\it for} }a\in V,
\end{eqnarray}
{\it is a $V$-homomorphism.}

{\sl Proof.} It follows from the proof of Proposition 3.4 [L1].$\;\;\;\;\Box$

For any complex numbers $c$ and $h$, let $M(c,h)$ be the Verma module
for the Virasoro algebra $Vir$ with central charge $c$ and with
lowest weight $h$. Let ${\bf 1}$ be a lowest weight vector of $M(c,0)$.
Then $L(-1){\bf 1}$ is a singular vector, i.e., $L(n){\bf 1}=0$ for
$n\ge 1$. Set $\bar{M}(c,0)=M(c,0)/<L(-1){\bf 1}>$, where $<L(-1){\bf
1}>$ denotes the submodule of $M(c,0)$ generated by $L(-1){\bf 1}$. Denote by
$L(c,h)$ the (unique) irreducible quotient module of $M(c,h)$. By
slightly abusing notations, we still use ${\bf 1}$ for the image of
${\bf 1}$ for both $\bar{M}(c,0)$ and $L(c,0)$.

{\bf Proposition 4.1.2}. {\it For any complex number $c$,
$\bar{M}(c,0)$ has a natural vertex operator algebra structure and any
restricted $Vir$-module $M$ of central charge $c$ is a weak
$\bar{M}(c,0)$-module. In
particular, for any complex number $h$, $M(c,h)$ is a $\bar{M}(c,0)$-module.}

{\sl Proof.}
Let $M$ be any restricted $Vir$-module with central charge $c$. Then
$\bar{M}(c,0)\oplus M$ is a restricted $Vir$-module. By Lemma 3.2.6,
$L(z)$ is an even local vertex operator on $(\bar{M}(c,0)\oplus M,L(-1))$.
Then by Corollary 3.2.11, $V=<L(z)>$ is a vertex algebra
with $\bar{M}(c,0)\oplus M$ as a module. Consequently, both
$\bar{M}(c,0)$ and $M$ are $V$-modules. By Lemma 2.3.5, the components
of $Y(L(z),z_{0})$ on $V$ satisfy the Virasoro relation. Since
$L(z)_{n}I(z)=0$ for $n\ge 0$,
$V$ is a lowest weight $Vir$-module with lowest weight $0$, so that $V$
is a quotient module of $\bar{M}(c,0)$. Let ${\bf 1}$ be
a lowest weight vector of $\bar{M}(c,0)$. Since
$L(-1){\bf 1}=0$, by Proposition 4.1.1, we have a $V$-homomorphism from $V$ to
$\bar{M}(c,0)$ mapping $I(z)$ to ${\bf 1}$. Then it follows that $V$
is isomorphic to $\bar{M}(c,0)$. Therefore $\bar{M}(c,0)$ is a
vertex operator algebra and any restricted $Vir$-module $M$ is a weak module.
$\;\;\;\;\Box$

{\bf Remark 4.1.3.} It follows that $L(c,0)$ is a quotient vertex
operator algebra of $\bar{M}(c,0)$. It follows from Kac determinant
formula that $L(c,0)=\bar{M}(c,0)$ for generic $c$. If $c$ is among a
certain discrete series of complex numbers, $L(c,0)\ne \bar{M}(c,0)$.
In this case, there are only finitely many irreducible
$L(c,0)$-modules (up to equivalence) [Wang].

\subsection{Vertex operator superalgebras and modules associated to
the Neveu-Schwarz algebra}
Let us first recall the definition of the Neveu-Schwarz algebra (cf.
 [FFR], [KW], [T]). The Neveu-Schwarz algebra is the Lie superalgebra
\begin{eqnarray}
NS=\oplus_{m\in {\bf Z}}{\bf C}L(m)\oplus \oplus_{n\in {\bf Z}}{\bf
C}G(n+\frac{1}{2})\oplus {\bf C}c
\end{eqnarray}
with the following commutation relations:
\begin{eqnarray}
& &[L(m),L(n)]=(m-n)L(m+n)+\frac{m^{3}-m}{12}\delta_{m+n,0}c,\\
& &[L(m),G(n+\frac{1}{2})]=\left(\frac{m}{2}-n-\frac{1}{2}\right)G(m+n+
\frac{1}{2}),\\
& &[G(m+\frac{1}{2}),G(n-\frac{1}{2})]_{+}=2L(m+n)+\frac{1}{3}m(m+1)
\delta_{m+n,0}c,\\
& &[L(m),c]=0,\;[G(n+\frac{1}{2}),c]=0.
\end{eqnarray}
By definition, $NS^{0}=\oplus_{m\in {\bf Z}}{\bf C}L(m)\oplus {\bf C}c$ and
$NS^{1}= \oplus_{n\in {\bf Z}}{\bf C}G(n+\frac{1}{2})$. We define
\begin{eqnarray}
\deg L(m)=m,\;\deg c=0,\;\deg G(n)=n\;\;\;\mbox{for }m\in {Z},n\in
{1\over 2}+{\bf Z}.
\end{eqnarray}
Then $NS=\oplus_{n\in {1\over 2}{\bf Z}}NS_{n}$ becomes a ${1\over
2}{\bf Z}$-graded Lie superalgebra. Set
\begin{eqnarray}
NS_{\pm}=\sum_{n=1}^{\infty}({\bf C}L(\pm n)+{\bf C}G(\pm n\mp{1\over 2})),\;
NS_{0}={\bf C}L(0)\oplus {\bf C}c.
\end{eqnarray}
Then we have the triangular decomposition $NS=NS_{+}\oplus
NS_{0}\oplus NS_{+}$.

Set
\begin{eqnarray}
L(z)=\sum_{m\in {\bf Z}}L(m)z^{-m-2},\;G(z)=\sum_{n\in {\bf
Z}}G(n+\frac{1}{2}) z^{-n-2}.
\end{eqnarray}
Then we have:
\begin{eqnarray}
& &[L(z_{1}),L(z_{2})]=z_{1}^{-1}\delta\left(\frac{z_{2}}{z_{1}}\right)
L'(z_{2})+2z_{1}^{-2}\delta'\left(\frac{z_{2}}{z_{1}}\right)L(z_{2})
+\frac{\ell}{12}z_{1}^{-4}\delta^{(3)}\left(\frac{z_{2}}{z_{1}}\right),\\
& &[L(z_{1}),G(z_{2})]\nonumber\\
&=&\sum_{m,n\in {\bf Z}}\left(\frac{m}{2}-n-\frac{1}{2}
\right)G(m+n+\frac{1}{2})z_{1}^{-m-2}z_{2}^{-n-2}\nonumber\\
&=&\sum_{m,n\in {\bf Z}}(-m-n-2)G(m+n+\frac{1}{2})z_{2}^{-m-n-3}z_{1}^{-1}
\left(\frac{z_{2}}{z_{1}}\right)^{m+1}\nonumber\\
& &+\sum_{m,n\in {\bf Z}}\frac{3}{2}(m+1)
G(m+n+\frac{1}{2})z_{2}^{-m-n-2}z_{1}^{-2}\left(\frac{z_{2}}{z_{1}}\right)^{m}
\nonumber\\
&=&z_{1}^{-1}\delta\left(\frac{z_{2}}{z_{1}}\right){\partial
\over\partial z_{2}}G(z_{2})+\frac{3}{2}\left({\partial\over\partial z_{2}}
z_{1}^{-1}\delta\left(\frac{z_{2}}{z_{1}}\right)\right)G(z_{2})
\end{eqnarray}
and
\begin{eqnarray}
& &[G(z_{1}),G(z_{2})]_{+}\nonumber\\
&=&\sum_{m,n\in {\bf Z}}[G(m+\frac{1}{2}),G(n-\frac{1}{2})]_{+}z_{1}^{-m-2}
z_{2}^{-n-1}\nonumber\\
&=&\sum_{m,n\in {\bf Z}}2L(m+n)z_{2}^{-m-n-2}z_{1}^{-1}\left(\frac{z_{2}}
{z_{1}}\right)^{m+1}+\sum_{m\in {\bf Z}}\frac{1}{3}m(m+1)cz_{1}^{-m-2}
z_{2}^{m-1}\nonumber\\
&=2&z_{1}^{-1}\delta\left(\frac{z_{2}}{z_{1}}\right)L(z_{2})+\frac{1}{3}c
\left({\partial\over\partial z_{2}}\right)^{2}z_{1}^{-1}\delta\left(\frac
{z_{2}}{z_{1}}\right).
\end{eqnarray}
It is clear that (4.2.2)-(4.2.4) are equivalent to (4.2.9)-(4.2.11),
respectively.

For any complex numbers $c$ and $h$, let $M_{c,h}$ be the Verma module
over $NS$ with lowest weight $h$ and with central charge $c$.
Then there exists a unique maximal proper submodule
$J_{c,h}$ of $M_{c,h}$. Denote the quotient module $M_{c,h}/J_{c,h}$
by $L_{c,h}$.  Recall that $v\in M_{c,h}$ is called a singular vector
if $NS_{+}v=0$ and $v$ is an eigenvector of $L(0)$. It is easy to see
that $G(-\frac{1}{2}){\bf 1}$ is a singular vector of $M_{c,0}$ for any $c$.
Denote the quotient module $M_{c,0}/\langle G(-\frac{1}{2}){\bf 1}\rangle$ by
$M_{c}$, where $\langle G(-\frac{1}{2})1\rangle$ is the submodule of
$M_{c,0}$ generated by the singular vector $G(-\frac{1}{2}){\bf 1}$.

Let $M$ be any restricted module with a central charge $c$ over the
Neveu-Schwarz algebra $NS$. Then  $L(z), G(z)\in  F(M)$.
By Lemma 2.1.3, we obtain
\begin{eqnarray}
(z_{1}-z_{2})^{2}[L(z_{1}),G(z_{2})]&=&0\\
(z_{1}-z_{2})^{3}[G(z_{1}),G(z_{2})]_{+}&=&0.
\end{eqnarray}
Then $\{L(z),G(z)\}$ is a set of mutually local homogeneous vertex operators on
$(M,L(-1))$. By Corollary 3.2.11, $V=<L(z),G(z)>$ is a vertex superalgebra with
$M$ as a $V$-module. By the same argument as one in proof of
Proposition 4.1.2, we obtain

{\bf Proposition 4.2.1.} {\it For any complex number $c$, $M_{c}$ has a
natural vertex operator superalgebra structure such that any
restricted $NS$-module $M$ with central charge $c$ is a weak $M_{c}$-module.}

\subsection{Vertex operator superalgebras associated to an affine
 Lie superalgebra}
Let $({\bf g},B)$ be a pair consisting of a finite-dimensional
Lie superalgebra ${\bf g}={\bf g}_{0}\oplus {\bf g}_{1}$ such that
$[{\bf g}_{1},{\bf g}_{1}]_{+}=0$ and a
nondegenerate symmetric bilinear form $B$ such that
\begin{eqnarray}
B({\bf g}_{0},{\bf g}_{1})=0,\; B([a,u],v)=-B(u,[a,v])\;\;\;\mbox{for }a\in
{\bf g}_{0}, u,v\in {\bf g}.
\end{eqnarray}
This amounts to having a finite-dimensional Lie algebra ${\bf g}_{0}$
with a nondegenerate symmetric invariant bilinear form
$B_{0}(\cdot,\cdot)$ and a finite-dimensional ${\bf g}_{0}$-module
${\bf g}_{1}$ with a nondegenerate symmetric bilinear form
$B_{1}(\cdot,\cdot)$ such that
\begin{eqnarray}
B_{1}(au,v)=-B(u,av)\;\;\;\mbox{ for any }a\in {\bf g}_{0}, u,v\in
{\bf g}_{1}.
\end{eqnarray}
Set
\begin{eqnarray}
\tilde{\bf g}={\bf C}[t,t^{-1}]\otimes {\bf g}\oplus {\bf C}c.
\end{eqnarray}
Then we define
\begin{eqnarray}
& &[a_{m},b_{n}]=[a,b]_{m+n}+m\delta_{m+n,0}\langle a,b\rangle c,\\
& &[a_{m},u_{n}]=-[u_{n},a_{m}]=(au)_{m+n},\\
& &[u_{m},v_{n}]_{+}=\delta_{m+n+1,0}\langle u,v\rangle c,\\
& &[c, x_{m}]=0
\end{eqnarray}
for any $a,b\in {\bf g}_{0}, u,v\in {\bf g}_{1}, x\in {\bf g}$ and
$m,n\in {\bf Z}$, where $x_{m}$ stands for $t^{m}\otimes x$. It is
easy to check that we obtain a Lie superalgebra $\tilde{{\bf g}}$ with
\begin{eqnarray}
\tilde{{\bf g}}_{0}={\bf C}[t,t^{-1}]\otimes {\bf g}_{0}\oplus {\bf C},\;\;
\tilde{{\bf g}}_{1}={\bf C}[t,t^{-1}]\otimes {\bf g}_{1}.
\end{eqnarray}
For any $x\in {\bf g}$, we set
\begin{eqnarray}
x(z)=\sum_{n\in {\bf Z}}x_{n}z^{-n-1}.
\end{eqnarray}
Then the defining relations (4.3.4)-(4.3.7) of $\tilde{{\bf g}}$ are
equivalent to the following equations:
\begin{eqnarray}
& &[a(z_{1}),b(z_{2})]=z_{1}^{-1}\delta\left(\frac{z_{2}}{z_{1}}\right)
[a,b](z_{2})+
z_{1}^{-2}\delta'\left(\frac{z_{2}}{z_{1}}\right)\langle
a,b\rangle c,\\
& &[a(z_{1}),u(z_{2})]=z_{1}^{-1}\delta\left(\frac{z_{2}}{z_{1}}\right)
(au)(z_{2}),\\
& &[u(z_{1}),v(z_{2})]_{+}=\langle u,v\rangle
z_{1}^{-1}\delta\left(\frac{z_{2}}{z_{1}}\right)c,\\
& &[x(z),c]=0
\end{eqnarray}
for any $a,b\in {\bf g}_{0}, u,v\in {\bf g}_{1}, x\in {\bf g}, m,n\in
{\bf Z}$.

Define
\begin{eqnarray}
\deg a_{m}=-m, \;\deg u_{n}=-n+{1\over 2},\; \deg c=0
\end{eqnarray}
for any $a\in {\bf g}_{0}, u\in {\bf g}_{1}, m,n\in {\bf Z}$. Then
$\tilde{{\bf g}}$ is a ${1\over 2}{\bf Z}$-graded Lie superalgebra.
Set
\begin{eqnarray}
N_{+}=t{\bf C}[t]\otimes {\bf g}_{0}\oplus {\bf
C}[t]\otimes {\bf g}_{1},\; N_{-}=t^{-1}{\bf C}[t^{-1}]\otimes {\bf
g},\;N_{0}={\bf g}_{0}\oplus {\bf C}c.
\end{eqnarray}
Then we obtain a triangular decomposition $\tilde{{\bf g}}=N_{+}\oplus
N_{0}\oplus N_{-}$. Let $P=N_{+}+N_{0}$ be the parabolic subalgebra.
For any ${\bf g}_{0}$-module $U$ and any complex
number $\ell$, denote by $M_{({\bf g},B)}(\ell,U)$ the generalized Verma
module or Weyl
module with $c$ acting as scalar $\ell$. Namely,
$M_{({\bf g},B)}(\ell,U)=U(\tilde{{\bf g}})\otimes_{U(P)}U$.
For any $\tilde{{\bf g}}$-module $M$, we may consider $x(z)$ for $x\in
{\bf g}$ as an element of $({\rm End}M)[[z,z^{-1}]]$. Recall that a
$\tilde{{\bf g}}$-module $M$ is said to be {\it restricted} if for any
$u\in M$, $(t^{k}{\bf C}[t]\otimes {\bf g})u=0$ for $k$ sufficiently
large. Then a $\tilde{{\bf g}}$-module $M$
is restricted if and only if $x(z)$ for all $x\in {\bf g}$ are weak
vertex operators on $M$.

{\bf Theorem 4.3.1.} {\it For any complex number $\ell$, $M_{({\bf
g},B)}(\ell,{\bf C})$ has a natural vertex superalgebra structure and any
restricted $\tilde{{\bf g}}$-module $M$ of level $\ell$ is a $M_{({\bf
g},B)}(\ell,{\bf C})$-module.}

{\sl Proof.} Let $M$ be any restricted $\tilde{{\bf g}}$-module of
level $\ell$. Then $W=M_{({\bf g},B)}(\ell,{\bf C})\oplus M$ is also a
restricted  $\tilde{{\bf g}}$-module of level $\ell$.
It follows from  Lemma 2.1.3 and (4.3.10)-(4.3.12) that
$\bar{{\bf g}}=\{a(z)|a\in {\bf g}\}$ is a local subspace of
$F(W)$. Let $V$ be the subspace of $F(W)$ generated by
all $\bar{\bf g}$. Then by Corollary 3.2.11, $V$ is a vertex
superalgebra and $W$ is a $V$-module. Consequently, both $M$ and
$M_{({\bf g},B)}(\ell,{\bf C})$ are $V$-modules.
It follows from Lemma 2.3.5 and
(4.3.10)-(4.3.12) that $V$ is a $\tilde{{\bf g}}$-module (of
level $\ell$) with a vector $I(z)$ satisfying $P\cdot I(z)=0$, so that $V$ is a
quotient $\tilde{{\bf g}}$-module of $M_{({\bf g},B)}(\ell,{\bf C})$.

To finish the proof, we only need to prove that $V$ is isomorphic to
$M_{({\bf g},B)}(\ell,{\bf C})$ as a $V$-module. Let $d$ be the
endomorphism of $M_{({\bf g},B)}(\ell,{\bf C})$ such
that
\begin{eqnarray}
d\cdot {\bf 1}=0,\;[d, a_{m}]=-ma_{m-1}\;\;\mbox{ for }a\in {\bf g}.
\end{eqnarray}
Then $[d,a(z)]=a'(z)$ for any $a\in {\bf g}$. Then $(M_{({\bf
g},B)}(\ell,{\bf C}),d)$ is a $V$-module. It
follows from Proposition 4.1.1 and the universal property of $M_{({\bf
g},B)}(\ell,{\bf C})$ that $V$ and $M_{({\bf g},B)}(\ell,{\bf C})$ are
isomorphic $V$-modules. $\;\;\;\;\Box$

{\bf Remark 4.3.2.} It is clear that $M_{({\bf
g},B)}(\ell,{\bf C})=M_{({\bf g},\alpha B)}(\alpha^{-1}\ell,{\bf C})$ for any
nonzero complex number $\alpha$ (see for example [Lian]).

In order to study $M_{({\bf g},B)}(\ell,{\bf C})$ more closely, we
first consider the following interesting special cases.

{\bf Case 1.} Let ${\bf g}_{0}$ be a finite-dimensional Lie algebra
with a fixed nondegenerate symmetric invariant bilinear form
$B(\cdot,\cdot)$ and let ${\bf g}_{1}=0$. Then for any
complex number $\ell$,  by Theorem 4.3.1 we
obtain a vertex algebra $M_{({\bf g},B)}(\ell,{\bf C})$.

Next, we shall show that if ${\bf g}$ is simple, $M_{({\bf
g},B)}(\ell,{\bf C})$ is a vertex operator algebra except for one $\ell$.
Let ${\bf g}$ be a finite-dimensional simple Lie algebra
with a fixed
Cartan subalgebra ${\bf h}$. Let $\Delta$ be the set of all roots, let
$\Pi $ be a set of simple roots, and let $\theta$ be the
highest root. Let $\langle\cdot,\cdot\rangle$ be the
normalized Killing form such that $\langle\theta,\theta\rangle=2$.
Then the Killing form on ${\bf g}$ is
$2\Omega \langle\cdot,\cdot\rangle$, where $\Omega$ is the dual Coxeter number.
Let $u^{i}$ $(i=1,\cdots,d)$ be an orthonormal
basis for ${\bf g}$.  Set
\begin{eqnarray}
\omega=\frac{1}{2(\Omega +\ell)}\sum_{i=1}^{d}u^{i}(-1)u^{i}(-1){\bf
1}\in M_{{\bf g}}(\ell,{\bf C}).
\end{eqnarray}
Since we have already proved that $M_{{\bf g}}(\ell,{\bf C})$ has a vertex
algebra structure, by a calculation in [DL] or [Lian], we have:

{\bf Proposition 4.3.3}. {\it For any complex number $\ell\ne
-\Omega$, $M_{{\bf
g}}(\ell,{\bf C})$ is a vertex operator algebra of rank
$\displaystyle{\frac{d\ell}{\Omega+\ell}}$ with the Virasoro element
$\omega$ given by (4.3.17) and any restricted $\tilde{{\bf
g}}$-module $M$ of level $\ell$ is a weak module for $M_{{\bf
g}}(\ell,{\bf C})$ being a VOA. In particular, for any finite-dimensional
${\bf g}$-module $U$, $M_{{\bf g}}(\ell,U)$ is a module for
$M_{{\bf g}}(\ell,{\bf C})$ being a VOA.}

{\bf Remark 4.3.4.} If $\ell = -\Omega$ (which is called the critical level),
then $Y(\omega,z)$ commutes with  $a(z)$ for any $a\in {\bf g}$. A
beautiful application of vertex algebra theory can be found in [FF]
where it was proved that the center of $M(-\Omega,{\bf C})$ is linearly spanned
by all lowest weight vectors (or singular vectors).

{\bf Case 2.} Let ${\bf g}_{1}$ be a $n$-dimensional vector space with
a nondegenerate symmetric bilinear form $\langle \cdot,\cdot\rangle$
and let ${\bf g}_{0}=0$. Then by Theorem 4.3.1, $M_{{\bf
g}_{1}}(1,{\bf C})$ is a vertex superalgebra. Since the vertex
superalgebra $M_{{\bf g}_{1}}(1,{\bf C})$ only depends on the positive
integer $n$ (up to isomorphism), we
set ${\bf F}^{n}=M_{{\bf g}_{1}}(1,{\bf C})$. Since ${\bf F}^{n}$ is
an irreducible module for the Clifford algebra $U(\tilde{{\bf
g}}_{1})$, ${\bf F}^{n}$ is a simple vertex superalgebra.
Let $\{u^{1},\cdots,u^{n}\}$ be an orthonormal basis for ${\bf g}_{1}$
and set
\begin{eqnarray}
\omega ={1\over 2}\sum_{i=1}^{n}u^{i}_{-2}u^{i}_{-1}{\bf 1}.
\end{eqnarray}
Then for any $u\in {\bf g}_{1}$, we have:
\begin{eqnarray}
& &u_{0}\omega ={1\over 2}\sum_{i=1}^{n}u_{0}u^{i}_{-2}u^{i}_{-1}{\bf 1}
=-{1\over 2}\sum_{i=1}^{n}\langle u,u^{i}\rangle u^{i}_{-2}{\bf 1}
=-{1\over 2}u_{-2}{\bf 1}
=-{1\over 2}Du,\\
& &u_{1}\omega ={1\over 2}\sum_{i=1}^{n}u_{1}u^{i}_{-2}u^{i}_{-1}{\bf 1}
={1\over 2}\sum_{i=1}^{n}\langle u,u^{i}\rangle u^{i}_{-1}{\bf 1}
={1\over 2}u
\end{eqnarray}
and
\begin{eqnarray}
u_{k}\omega =0\;\;\mbox{ for any }k\ge 2.
\end{eqnarray}
By the commutator formula (2.2.6), we have:
\begin{eqnarray}
[\omega_{m+1},
u_{n}]=-[u_{n},\omega_{m+1}]=-\left(n+\frac{m+1}{2}\right)u_{m+n}
\end{eqnarray}
for any $m,n\in {\bf Z}$.

{\bf Proposition 4.3.5.} {\it ${\bf F}^{n}$ is a vertex operator
superalgebra of rank $\displaystyle{{n\over 2}}$ with the Virasoro
element $\omega$
given by (4.3.18).
Furthermore, any lower truncated ${1\over 2}{\bf Z}$-graded ${\bf
F}^{n}$-module is a direct sum of copies of adjoint module ${\bf
F}^{n}$, i.e., ${\bf F}^{n}$ is rational in the sense of [Zhu].}

{\sl Proof.}  To prove that
$\omega$ is a Virasoro element with central charge
$\displaystyle{{n\over 2}}$, By Lemma 2.3.5 it is equivalent to check
the following conditions:
\begin{eqnarray}
& &Y(\omega_{0}a,z)={d\over dz}Y(a,z)\;\;\mbox{for any }a\in {\bf F}^{n},\\
& &\omega _{1}\omega =2\omega,\;\omega_{2}\omega =0,\; \omega_{3}\omega
=\frac{n}{4},\;\omega_{m}\omega=0\;\;\mbox{for }m\ge 4.
\end{eqnarray}
These can be easily obtained by using (4.3.22) and (4.3.18). Let $M$
be any lower truncated ${1\over 2}{\bf Z}$-graded ${\bf F}^{n}$-module.
Then for any $u\in M$, there is a positive integer $k$ such that
\begin{eqnarray}
(v^{1})_{n_{1}}(v^{2})_{n_{2}}\cdots (v^{m})_{n_{m}}u=0\;\;\;\mbox{if
}n_{1}+\cdots+ n_{m}>k\;\;\mbox{for any }v^{i}\in {\bf g}_{1}.
\end{eqnarray}
Then the complete reducibility follows from a standard theorem.
$\;\;\;\;\Box$

{\bf Case 3.} Let ${\bf g}={\bf g}_{0}\oplus {\bf g}_{1}$ where ${\bf
g}_{0}$ is a finite-dimensional simple Lie
algebra with the normalized Killing form
$\langle\cdot,\cdot\rangle_{0}$ as in Case 1 and ${\bf
g}_{1}$ is a finite-dimensional ${\bf g}_{0}$-module with a nondegenerate
symmetric bilinear form $\langle\cdot,\cdot\rangle_{1}$ such that
$\langle au,v\rangle=-\langle u,av\rangle$ for any $a\in {\bf g}_{0},
u,v\in {\bf g}_{1}$. Set
$B(\cdot,\cdot)=\langle\cdot,\cdot\rangle_{0}\oplus
\langle\cdot,\cdot\rangle_{1}$.

Let $V^{1}$ and $V^{2}$ be two vertex superalgebras. Then it is well
known ([B], [FHL]) that $V^{1}\otimes V^{2}$ has a vertex superalgebra
structure. We shall prove that if $\ell \ne 0$, $M_{({\bf
g},B)}(\ell,{\bf C})$ is
isomorphic to the tensor product vertex superalgebra of a vertex algebra
$M_{({\bf g}_{0},B')}(1,{\bf C})$ with the vertex operator
superalgebra ${\bf F}^{n}$, where $B'$ is a certain nondegenerate
symmetric invariant bilinear form on ${\bf g}_{0}$.

{\bf Proposition 4.3.6.} {\it Let $V$ be a vertex superalgebra
and let
$V^{1}$ and $V^{2}$ be two vertex subsuperalgebras of $V$ such that
$V^{1}$ and $V^{2}$ generate $V$ and that}
\begin{eqnarray}
[Y(u^{1},z_{1}),Y(u^{2},z_{2})]=0\;\;\;\mbox{{\it for any }}u^{i}\in V^{i}.
\end{eqnarray}
{\it Suppose that $V^{2}$ is a vertex operator superalgebra itself
such that $\ker_{V^{2}} D={\bf C}{\bf 1}$. Then
$V$ is isomorphic to the tensor product vertex superalgebra}
$V^{1}\otimes V^{2}$.

{\sl Proof.} Define a linear map $\psi$ from $V^{1}\otimes V^{2}$ to $V$
such that $\psi (a\otimes u)=a_{-1}u$ ($=u_{-1}a$) for $a\in
V^{1},u\in V^{2}$. For any $a,b\in V^{1}, u,v\in V^{2}$, we have:
\begin{eqnarray}
\psi(Y(a\otimes {\bf 1},z)(b\otimes u))
&=&\psi(Y(a,z)b\otimes Y({\bf 1},z)u)\nonumber\\
&=&(Y(a,z)b)_{-1}u\nonumber\\
&=&u_{-1}Y(a,z)b\nonumber\\
&=&Y(a,z)u_{-1}b\nonumber\\
&=&Y(a,z)b_{-1}u\nonumber\\
&=&Y(\psi(a\otimes {\bf 1}),z)\psi(b\otimes u)\nonumber\\
&=&Y(\psi(a\otimes {\bf 1}),z)\psi(b\otimes u).
\end{eqnarray}
Similarly, we have:
\begin{eqnarray}
\psi(Y({\bf 1}\otimes u,z)(b\otimes v))=Y(\psi({\bf 1})\otimes
u,z)\psi(b\otimes v).
\end{eqnarray}
Since ${\bf 1}\otimes V^{2}$ and $V^{1}\otimes {\bf 1}$ generate
$V^{1}\otimes V^{2}$, $\psi$ is a vertex superalgebra homomorphism. Since
$V^{1}$ and $V^{2}$ generate $V$, $\psi$ is surjective. Since
$\ker_{V^{2}} L(-1)={\bf C}{\bf 1}$, it is easy to see that $\ker
({\bf 1}\otimes
L(-1))=V^{1}\otimes {\bf 1}$. Suppose  $\ker \psi\ne 0$. Since $V^{1}\otimes
V^{2}$ as a $V^{2}$-module is a direct sum of copies of $V^{2}$'s,
$\ker \psi$ contains at least one copiy of $V^{2}$, so that there
is a nonzero vector $a\in V^{1}$ such that $a\otimes {\bf 1}\in \ker
\psi$. Thus $a=a_{-1}{\bf 1}=\psi(a\otimes {\bf 1})=0$.  This is a
contradiction. Therefore, $\psi$ is injective. That is, $\psi$ is an
isomorphism. $\;\;\;\;\Box$

By Theorem 4.3.1, for any complex number $\ell$, $M_{({\bf
g},B)}(\ell,{\bf C})$ is a vertex superalgebra. We may consider
${\bf g}$ as a subspace of $M_{({\bf g},B)}(\ell,{\bf C})$ by
identifying $a$ with $a_{-1}{\bf 1}$ for any $a\in {\bf g}$.

Let $\{u^{1},\cdots,u^{n}\}$
be an orthonormal basis for ${\bf g}_{1}$. For any $a,b\in {\bf
g}_{0}$, we define
\begin{eqnarray}
B_{2}(a,b)=\sum_{i=1}^{n}B(au^{i},bu^{i}).
\end{eqnarray}
Let $\rho$ be the representation of ${\bf g}_{0}$ on ${\bf g}_{1}$. Then
\begin{eqnarray}
B_{2}(a,b)=\sum_{i=1}^{n}B(au^{i},bu^{i})=-\sum_{i=1}^{n}B(u^{i},abu^{i})=-
tr_{{\bf g}_{0}}\rho (a)\rho (b).
\end{eqnarray}
Thus $B_{2}$ is a symmetric invariant bilinear form on ${\bf g}_{0}$. Set
$B'=B+B_{2}$.

For any $u\in {\bf g}_{1}$, we have
$\displaystyle{u=\sum_{i=1}^{n}\langle u,u^{i}\rangle u^{i}}$.
Therefore
\begin{eqnarray}
a\cdot u^{i}=\sum_{i=1}^{n}\langle au^{i},u^{j}\rangle
u^{j}\;\;\mbox{for any }a\in {\bf g}_{0}.
\end{eqnarray}
For any $a\in {\bf g}_{0}$, set
\begin{eqnarray}
\bar{a}=a+{1\over
2\ell}\sum_{i=1}^{n}u^{i}_{-1}(au^{i})_{-1}{\bf 1}
\left(=a+{1\over 2\ell}\sum_{i,j=1}^{n}\langle au^{i},u^{j}\rangle
u^{i}_{-1}u^{j}_{-1}{\bf 1}\right).
\end{eqnarray}
For any $a\in {\bf g}_{0}, u\in {\bf g}_{1}$, by definition we have:
\begin{eqnarray}
u_{0}\bar{a}&=&u_{0}a+{1\over 2\ell}\sum_{i=1}^{n}
u_{0}u^{i}_{-1}(au^{i})_{-1}{\bf 1}\nonumber\\
&=&u_{0}a+{1\over 2}\sum_{i=1}^{n}\left(\langle u,u^{i}\rangle au^{i}
-\langle u,au^{i}\rangle
u^{i}_{-1}{\bf 1}\right)\nonumber\\
&=&u_{0}a+{1\over 2}au+{1\over 2}\sum_{i=1}^{n}\langle au,u^{i}\rangle
u^{i}_{-1}{\bf 1}\nonumber\\
&=&u_{0}a+au\nonumber\\
&=&0
\end{eqnarray}
and
\begin{eqnarray}
u_{k}\bar{a}=u_{k}a+{1\over 2\ell}\sum_{i,j=1}^{n}\langle au^{i},u^{j}\rangle
u_{k}u^{i}_{-1}u^{j}_{-1}{\bf 1}=0\;\;\mbox{ for }k\ge 1.
\end{eqnarray}
Thus by the commutator formula (2.2.6) we get
\begin{eqnarray}
[u_{m}, \bar{a}_{n}]=0\;\;\;\mbox{ for any }m,n\in {\bf Z}.
\end{eqnarray}
Let $V^{1}$ and $V^{2}$ be the vertex subsuperalgebras of
 $M_{({\bf g},B)}(\ell,{\bf C})$
generated by $\bar{{\bf g}}_{0}=\{\bar{a}|a\in {\bf g}_{0}\}$ and by
${\bf g}_{1}$, respectively. Since $V^{1}$ is generated by even
elements, $V^{1}$ is a vertex algebra. It is clear that
$V^{2}\simeq {\bf F}^{n}$. By (4.3.32)
and the iterate formula (2.2.8), we get:
\begin{eqnarray}
[a_{m}, u_{n}]=0\;\;\;\mbox{for any }a\in V^{1},u\in V^{2}.
\end{eqnarray}
Since ${\bf F}^{n}$ is a simple vertex operator superalgebra, it
follows from the proof of Lemma 4.3 [L1] or [T] that $\ker L(-1)={|bf
C}{\bf 1}$.
By Proposition 4.3.6, $M_{({\bf g},B)}(\ell,{\bf C})$ is isomorphic to
the tensor
product vertex superalgebra of vertex algebra $V^{1}$ with the vertex
operator superalgebra ${\bf F}^{n}$.

Let $a,b\in {\bf g}_{0}$. Then  we have:
\begin{eqnarray}
\bar{a}_{0}\bar{b}&=&a_{0}\bar{b}+{1\over 2\ell}\sum_{i=1}^{n}
\left(u^{i}_{-1}(au^{i})\right)_{0}\bar{b}_{-1}{\bf 1}\nonumber\\
&=&a_{0}\bar{b}\nonumber\\
&=&a_{0}b+{1\over 2\ell}\sum_{i=1}^{n}
a_{0}u^{i}_{-1}(bu^{j})\nonumber\\
&=&a_{0}b+{1\over 2\ell}\sum_{i=1}^{n}
\left((au^{i})_{-1}(bu^{i})+u^{i}_{-1}(abu^{i})\right)\nonumber\\
&=&a_{0}b+{1\over 2\ell}\sum_{i,j=1}^{n}
\langle au^{i},u^{j}\rangle u^{j}_{-1}(bu^{i})+{1\over 2\ell}\sum_{i=1}^{n}
u^{i}_{-1}(abu^{i})\nonumber\\
&=&a_{0}b-{1\over 2\ell}\sum_{i,j=1}^{n}\langle u^{i},au^{j}\rangle
u^{j}_{-1}(bu^{i})+{1\over 2\ell}\sum_{i=1}^{n}
u^{i}_{-1}(abu^{i})\nonumber\\
&=&a_{0}b-{1\over 2\ell}\sum_{j=1}^{n}u^{j}_{-1}(bau^{j})
+{1\over 2\ell}\sum_{i=1}^{n}
u^{i}_{-1}(abu^{i})\nonumber\\
&=&a_{0}b+{1\over 2\ell}\sum_{i=1}^{n}
u^{i}_{-1}([a,b]u^{i})\nonumber\\
&=&\overline{[a,b]},
\end{eqnarray}
and
\begin{eqnarray}
\bar{a}_{1}\bar{b}&=&a_{1}\bar{b}+{1\over 2\ell}\sum_{i=1}^{n}\left(u^{i}_{-1}
au^{i}\right)_{1}\bar{b}\nonumber\\
&=&a_{1}\bar{b}\nonumber\\
&=&a_{1}b+{1\over 2\ell}\sum_{i=1}^{n}a_{1}u^{i}_{-1}(bu^{i})\nonumber\\
&=&\left(\ell B(a,b)+{1\over 2}B_{2}(a,b)\right){\bf 1}.
\end{eqnarray}
It is easy to see that $a_{n}\bar{b}=0$ for $n\ge 2$. Thus
\begin{eqnarray}
& &\bar{a}_{n}\bar{b}=\bar{a}_{n}b=0\;\;\;\mbox{ for }n\ge 2.
\end{eqnarray}
By the commutator formula (2.2.6), we have:
\begin{eqnarray}
[\bar{a}_{m},\bar{b}_{n}]=\overline{[a,b]}_{m+n}+B'(a,b)
\end{eqnarray}
for any $a,b\in {\bf g}_{0}, m,n\in {\bf Z}$. Therefore, $\bar{{\bf
g}}_{0}=\{\bar{a}|a\in {\bf g}_{0}\}$ is a Lie algebra isomorphic to ${\bf
g}_{0}$ under the multiplication $[\bar{a},\bar{b}]=\overline{[a,b]}$.
Then we have $V^{1}\simeq M_{({\bf g}_{0},B')}(\ell,{\bf C})$.
Therefore we obtain:

{\bf Proposition 4.3.7.} {\it Let $({\bf g},B)$ be any pair
satisfying (4.3.1) and let $\ell$ be any
nonzero complex number. Then vertex superalgebra $M_{({\bf
g},B)}(\ell,{\bf C})$
is isomorphic to the tensor vertex superalgebra $M_{({\bf
g}_{0}, B')}(1,{\bf C})\otimes {\bf F}^{n}$.}

{\bf Corollary 4.3.8.} {\it For any complex number $\ell \ne 0$,
$M_{{\bf g}}(\ell,{\bf C})$ is a vertex operator superalgebra of rank
$\displaystyle{{n\over 2}+\frac{(\ell-\Omega)d}{\ell}}$.}

{\sl Proof.} Since $B_{2}=-2\Omega B$, we have $B'=(\ell-\Omega)B$.
Then this corollary follows from Propositions 4.3.3, 4.3.5 and 4.3.7.
$\;\;\;\;\Box$

Since $M_{({\bf g},B)}(\ell,{\bf C})$ as a vertex operator
superalgebra is generated by
${\bf g}$, it follows from the iterate formula (2.2.8) that the
(unique) maximal proper $\tilde{{\bf g}}$-submodule of $M_{({\bf
g},B)}(\ell,{\bf C})$ is a
module for the vertex superalgebra $M_{({\bf g},B)}(\ell,{\bf C})$.
Therefore, the uniquely determined irreducible quotient module
$L_{({\bf g},B)}(\ell,0)$ is a quotient vertex operator superalgebra.

{\bf Corollary 4.3.9.} {\it The vertex operator superalgebra $L_{{\bf
g}}(\ell,0)$ is isomorphic to the tensor product vertex operator
superalgebra of the
vertex operator algebra $L_{{\bf g}_{0}}(\ell-\Omega,0)$ with the
vertex operator superalgebra ${\bf F}^{n}$.}

{\bf Remark 4.3.10.} In [KW], the authors have studied the special case
when ${\bf g}_{0}$ is simple and ${\bf g}_{1}={\bf g}_{0}$ is the
adjoint module.

\newpage
\section{ The semisimple representation theory for $L_{{\bf g}}(\ell,0)$}
In this section we shall prove that if $\ell$ is a positive integer, then
any lower truncated ${\bf Z}$-graded weak $L_{{\bf g}}(\ell,0)$-module is
completely reducible and the set of equivalence classes of
irreducible $L_{{\bf g}}(\ell,0)$-modules is exactly the set of
equivalence classes of standard $\tilde{{\bf g}}$-modules of level $\ell$.
This in particular proves the rationality  of
$L_{{\bf g}}(\ell,0)$ (in the sense of [Zhu]).

\subsection{The semisimple quotient algebras of $U({\bf g})$}

In order to prove the rationality of $L_{{\bf g}}(\ell,0)$, we need a
result about complete reducibility for a class of ${\bf g}$-modules.
In this subsection we study the semi-simplicity for some quotient
algebras of the universal enveloping algebra $U({\bf g})$ of a given
simple Lie algebra ${\bf g}$.
Let ${\bf g}$ be a finite-dimensional simple Lie algebra with a fixed Cartan
subalgebra ${\bf h}$. Let $\Delta$ be the set of roots, and fix a set
$\Pi$ of simple roots. Let $\theta$ be
the highest root. For any positive integer $\ell$ and $\alpha\in
\Delta$, define $A_{{\bf g}}(\alpha,\ell)$
to be the quotient algebra of $U({\bf g})$ modulo the two-sided ideal
generated by ${\bf g}_{\alpha}^{\ell +1}$.

{\bf Proposition 5.1.1.} {\it Let ${\bf g}=sl(2,{\bf C})$ and let
$\ell$ be any positive integer. Then the quotient
algebra $A=U({\bf g})/\langle e^{\ell +1}\rangle$ is semisimple.
Equivalently, any $A$-module is completely reducible.}

{\sl Proof}. Step one: Any nonzero $A$-module $M$ contains an
irreducible submodule of dimension at most $\ell+1$.
Let $r$ be the nonnegative integer such that $e^{r}M\ne 0$ and $e^{r+1}M=0$.
If $r=0$, or equivalently $eM=0$, then $hM=[e,f]M=0$ and $fM={1\over
2}[f,h]M=0$. Therefore $M$ is a direct sum of one-dimensional modules.
Next, we assume $r\ne 0$. Since
\begin{eqnarray}
[e^{r+1},f]=(r+1)(h-r)he^{r}, \end{eqnarray}
we have
\begin{eqnarray}
eu=0,\; hu=ru\;\;\;\;\mbox{for any }u\in e^{r}M.\end{eqnarray}
Let $0\ne u\in e^{r}M$. Then $u$ generates a highest
weight module $M_{1}$ of highest weight $r$. If $u_{r}=f^{r+1}u\ne 0$, then
$u_{r}$ generates a highest weight module of highest weight $-r$, which is
infinite-dimensional and irreducible. Thus $e^{n}f^{n}u_{r}\ne 0$ for
any positive integer $n$. This is a contradiction. Therefore
$f^{r+1}u=0$. Consequently, $M_{1}$ is a $(r+1)$-dimensional
irreducible submodule.

Step two: Any $A$-module $M$ is completely reducible. Let $M^{1}$ be
the sum of all irreducible submodules of dimension at most $\ell+1$. If
$M=M^{1}$, then $M$ is completely reducible. Otherwise, by Step one,
there is a submodule $W$ of $M$ such that $W/M^{1}$ is an
irreducible module of dimension at most $\ell+1$. It is clear that
both $e$ and $f$ nilpotently act on $W$ and $h$ locally finitely acts
on $W$. Then $W$ is completely reducible as a ${\bf g}$-module. That
is, there is an irreducible submodule $M^{2}$ of dimensional at most
$\ell+1$ such that $W=M^{1}\oplus M^{2}$. This contradicts to the
choice of $M^{1}$. Therefore, $M=M^{1}$ is completely reducible.$\;\;\;\;\Box$

{\bf Proposition 5.1.2}. {\it Let $\theta$ be the highest root of
${\bf g}$. Then any $A_{{\bf g}}(\theta,\ell)$-module $M$ is a direct
sum of finite-dimensional irreducible ${\bf g}$-modules $L_{{\bf
g}}(\lambda)$ such that $\langle \lambda,\theta\rangle\le \ell$. Consequently,
 $A_{{\bf g}}(\theta,\ell )$ is semisimple.}

{\sl Proof.} For any positive root of $({\bf
g},{\bf h})$, we have a copy of $sl(2,{\bf C})$ inside ${\bf g}$,
denoted by ${\bf
g}^{\alpha}$ with a basis $\{e_{\alpha},f_{\alpha},h_{\alpha}\}$. It
follows from Proposition 5.1.1 that $M=\oplus_{n=-\ell}^{\ell}M_{n}$, where
$M_{n}$ is the eigenspace of eigenvalue $n$ for $h_{\theta}$. Since
\begin{eqnarray}
e_{\alpha}\cdot M_{n}\subseteq M_{n+\langle
\theta,\alpha\rangle}\;\;\;\mbox{ for any }n,\;\mbox{and }0\ne\langle \theta,
\alpha\rangle\in {\bf Z},\end{eqnarray}
we get $e_{\alpha}^{2\ell +2}M=0$. It follows from Proposition 5.1.1
that $M$ is
a direct sum of finite-dimensional ${\bf g}^{\alpha}$-modules. Let
$\alpha$ go through all positive roots. Then ${\bf h}$
acts semisimply on $M$. Since all $e_{\alpha},f_{\alpha}$ act
nilpotently on $M$,
$M$ is completely reducible as a ${\bf g}$-module. Since the highest
weight $\lambda$ of any irreducible ${\bf g}$-submodule is also a highest
weight for ${\bf g}^{\theta}$, then $\langle \lambda,
\theta\rangle\le \ell.\;\;\;\;\Box$

{\bf Remark 5.1.3.} From [FZ], $A_{{\bf g}}(\theta,\ell)$ is Zhu's algebra
 $A(L_{{\bf g}}(\ell,0))$ for the vertex operator algebra $L_{{\bf
g}}(\ell,0)$.
Furthermore, $A_{{\bf g}}(\theta,\ell)$ is closely related to the
quotient algebra of quantum group $U_{q}({\bf g})$ with $q={\rm
exp}\left(\frac{2\pi i}{\ell +2}\right)$ in a certain subtle way.

{\bf Remark 5.1.4.} Let $A$ be any
finite-dimensional quotient algebra of $U({\bf g})$. Since $A$ is a
finite-dimensional ${\bf g}$-module, there is a positive integer $\ell$
such that $e_{\theta}^{\ell +1}A=0$. Therefore $A$ is a quotient algebra
of $A({\bf g},\ell)$. Consequently, $A$ is semisimple.

{\bf Proposition 5.1.5.} {\it Let $\alpha$ be any root of ${\bf g}$ with
respect to ${\bf h}$, and let $k$ be any nonnegative integer. Then
the quotient algebra $A_{{\bf g}}(\alpha,\ell)$ of $U({\bf g})$
 is semisimple.}

{\sl Proof.} Let $M$ be any $A_{{\bf g}}(\alpha,\ell)$-module.
 As before, ${\bf g}^{\alpha}={\bf g}_{\alpha}+{\bf
C}h_{\alpha}+{\bf g}_{-\alpha}$ is a subalgebra isomorphic to
$sl(2,{\bf C})$. By Proposition 5.1.1, we have:
\begin{eqnarray}
M=\oplus_{i=-k}^{k}M(i)\;\;\;\mbox{where }M(i)=\{u\in M|h_{\alpha}u=iu\}.
\end{eqnarray}
Just as in the proof of Proposition 5.1.2, we obtain
${\bf g}_{\theta}^{2k+2}M=0$. In particular, if $M=A_{{\bf
g}}(\alpha,k)$, then this implies that $A_{{\bf g}}(\alpha,k)$ is a
quotient algebra of $A_{{\bf g}}(\theta,2k+2)$. It follows from
Proposition 5.1.2 and Remark 5.1.4 that $A_{{\bf g}}(\alpha,k)$ is semisimple.
$\;\;\;\;\Box$

{\bf Remark 5.1.6.} This section was written in the first preprint of
this paper about one year ago. Later, we noticed that
Kac and Wang [KW] gave a more general result by using a Lie group
approach which is much simpler.
Since we think that our algebraic approach may be useful to certain
quadratic algebras, we just keep as it was.

\subsection{Untwisted representation theory for $L_{{\bf g}}(\ell,0)$}

The following two propositions are well
known (at least for simple Lie algebras of type $A$, $D$ or $E$).
In this subsection we give a different proof without using the explicit
constructions of basic standard modules.

{\bf Proposition 5.2.1}. {\it Let $e_{\alpha}$ be any root vector of ${\bf g}$
with root $\alpha$. Suppose $L_{{\bf g}}(\ell,0)$ is an integrable
$\tilde{{\bf g}}$-module. Then
$Y(e_{\alpha},z)^{t\ell +1}=0$ acting on $L_{{\bf g}}(\ell,0)$ where $t=1$ if
$\alpha$
is a long root, $t=2$ if $\alpha$ is a short root and ${\bf
g}\ne {\bf G}_{2}$, $t=6$ if $\alpha$ is a short root and ${\bf
g}={\bf G}_{2}$.}

{\sl Proof}. By Dong and Lepowsky's Proposition 2.3.6, it is
equivalent to prove that
\begin{eqnarray}
(e_{\alpha})_{n}e_{\alpha}=0\;\;\mbox{ for all }n\in {\bf
Z}_{+};\;\;(e_{\alpha})_{-1}^{t\ell}e_{\alpha}=(e_{\alpha})_{-1}^{t\ell +1}{\bf
1}=0.
\end{eqnarray}
{}From the standard semisimple Lie algebra theory (cf. [Hum], [Jac], [Kac])
we can embed $sl(2,{\bf C})$ into ${\bf g}$
as ${\bf g}^{\alpha}$ linearly spanned by
$e_{\alpha},f_{\alpha},h_{\alpha}$. Since
$\langle h_{\alpha},h_{\alpha}\rangle =4\langle \alpha,\alpha
\rangle^{-1}=2t$, so $\tilde{sl}(2,{\bf
C})$ becomes a subalgebra of $\tilde{{\bf g}}$ with the central
element $tc$. Then
it follows from the standard results [Kac] for integrable highest weight
representation
theory for an affine Lie algebra.$\;\;\;\;\Box$

{\bf Proposition 5.2.2}. {\it Let ${\bf g}=sl(2,{\bf C})$. Then
$Y(e,z)^{\ell +1}=
Y(f,z)^{\ell +1}=0$ acting on
$L_{{\bf g}}(\ell ,m)$ for $0\le m\le \ell$.}

{\sl Proof}. By Proposition 5.2.1,
$Y(e,z)^{2}=Y(f,z)^{2}=0$ on
$L_{{\bf g}}(1,0)$. Let $\sigma$ be the involution of $\tilde{sl}(2,{\bf
C})$, defined as follows:
\begin{eqnarray}
\sigma &:&t^{n}\otimes e\mapsto t^{n+1}\otimes f,\;\;t^{n}\otimes
f\mapsto t^{n-1}\otimes e,\\
& &t^{n}\otimes h\mapsto \delta _{n,0}c-t^{n}\otimes h,\;\;c\mapsto
c\;\;\mbox{for }n\in {\bf Z}.
\end{eqnarray}
In terms of generating functions, $\sigma$ may be written as
\begin{eqnarray}
\sigma (Y(e,z))=zY(f,z),\;\sigma (Y(f,z))=z^{-1}Y(e,z),\;\sigma
(Y(h,z))=c-Y(h,z).
\end{eqnarray}
Then through $\sigma$, $L_{{\bf g}}(1,1)$ becomes a $\tilde{{\bf
g}}$-module which is isomorphic to $L_{{\bf g}}(1,0)$. Then
$Y(e,z)^{2}=Y(f,z)^{2}=0$ acting on $L_{{\bf g}}(1,1)$ as well.

For any positive integer $\ell$ and any $0\le m\le \ell$. Since
$L_{{\bf g}}(\ell,m)$ can be embedded into the $m$ tensor $L_{{\bf
g}}(\ell,0)^{\otimes (\ell -m)}\otimes L_{{\bf g}}(\ell,1)^{\otimes
m}$, we have
$$Y(e,z)^{\ell +1}=Y(f,z)^{\ell +1}=0\;\;\;\;(\mbox{acting on }L_{{\bf g}}(\ell
,m)).\;\;\;\;\Box$$

{\bf Proposition 5.2.3}. {\it Let $e$ be any root
vector of ${\bf g}$ with root $\alpha$. If $L_{{\bf g}}(\ell,\lambda)$
is an integrable $\tilde{{\bf g}}$-module, then
$Y(e,z)^{t\ell +1}=0$ acting on $L_{{\bf g}}(\ell,\lambda)$.}

{\sl Proof.} For any root $\alpha$ of $({\bf
g},{\bf h})$, we can embed $\tilde{sl}(2,{\bf C})$ into $\tilde{{\bf
g}}$ with  center $tc$. Then $L_{{\bf g}}(\ell,\lambda)$ is a direct
sum of standard $\tilde{sl}(2,{\bf C})$-modules of level $t\ell$. From
Proposition 5.2.2, we have:
\begin{eqnarray}
Y(e,z)^{t\ell +1}=0\;\;\;\;(\mbox{acting on }L_{{\bf
g}}(\ell,\lambda))\end{eqnarray}
for any vector $e\in {\bf g}_{\alpha}.\;\;\;\;\Box$

{\bf Proposition 5.2.4}. {\it Any standard $\tilde{{\bf g}}$-module $L_{{\bf
g}}(\ell,\lambda)$ is a module for the vertex operator algebra
$L_{{\bf g}}(\ell,0)$.}

{\sl Proof}. Let $V$ be the subspace of ${\rm F}(L_{{\bf g}}(\ell,0))$
generated by all $a(z)=\sum_{n\in {\bf Z}}a(n)z^{-n-1}$ for $a\in {\bf g}$. By
Corollary 3.2.11 and Proposition 4.3.3, $V$ is a vertex operator
algebra (which is a quotient module of $M_{{\bf g}}(\ell,0)$) and
$L_{{\bf g}}(\ell,0)$ is  a faithful module.
By Proposition 5.2.3, $Y(e_{\theta},z)^{\ell+1}=0$
on $L_{{\bf g}}(\ell,\lambda)$.
It follows from Remark 2.3.7 that $Y(e_{\theta},z)^{\ell+1}=0$ on
$V$. Since $Y(e,z)^{\ell +1}{\bf 1}=0$ implies that
$(e_{-1})^{\ell +1}{\bf 1}=0$ (the constant term), $V$ is an integrable
$\tilde{{\bf g}}$-module. Therefore $V=L_{{\bf g}}(\ell,0)$. That is,
$L_{{\bf g}}(\ell,\lambda)$ is a module for the vertex operator
algebra  $L_{{\bf g}}(\ell,0)$. $\;\;\;\;\Box$

{\bf Proposition 5.2.5}. {\it Let $M=\oplus_{n\in {\bf Z}}M(n)$ be
any lower truncated ${\bf Z}$-graded
weak $L_{{\bf g}}(\ell,0)$-module. Then $M$
is a direct sum of standard
$\tilde{{\bf g}}$-modules of level $\ell$.}

{\sl Proof.} Step one: Any nonzero  ${\bf Z}$-graded
weak $L_{{\bf g}}(\ell,0)$-module $M=\oplus_{n\in {\bf Z}} M(n)$
truncated from below  contains
some graded submodule which is a standard $\tilde{{\bf g}}$-module of
level $\ell$.
Let $n$ be the integer such that $M(n)\ne 0$ and
$M(m)=0$ for $m<n$. Then
\begin{eqnarray}
Y(e_{\theta},z)^{\ell+1}M(n)=0.\end{eqnarray}
Extracting the coefficient of $z^{-\ell-1}$ from (5.2.6), we obtain
$(e_{\theta})_{0}^{\ell+1}M(n)=0$.
{}From Proposition 5.1.2, $M(n)$ is a direct sum of finite-dimensional
${\bf g}$-modules $L(\lambda)$ such that $\langle\lambda, \theta\rangle\le
\ell$. Let
$u$ be any highest weight vector for ${\bf g}$ in $M(n)$. Extracting
the constant from $Y(e_{\theta},z)^{\ell +1}u=0$, we obtain
$(e_{\theta})_{-1}^{\ell +1}u=0$.
 Let $M_{1}$ be the
submodule generated by $u$. Then $M_{1}$ is an integrable lowest weight
$\tilde{{\bf g}}$-module (the semisimplicity of ${\bf h}$ is obvious). Then
$M_{1}$ is a standard $\tilde{{\bf g}}$-module.

Step two: Any lower truncated ${\bf Z}$-graded weak $L_{{\bf
g}}(\ell,0)$-module is completely reducible.
Let $M_{1}$ be the sum of all graded standard
$\tilde{{\bf g}}$-modules inside $M$. If $M\ne M_{1}$, by Step one,
$M/M_{1}$ contains a graded standard $\tilde{{\bf g}}$-module,
i.e., there is a submodule $M_{2}$ of
$M$ such that $M_{2}/M_{1}$ is a standard $\tilde{{\bf g}}$-module.
Since $M_{1}$
and $M^{2}/M^{1}$ are completely reducible ${\bf
g}$-module, so is $M_{2}$. Thus ${\bf h}$ acts semisimply on $M_{2}$.
Then $M_{2}$ is an integrable $\tilde{{\bf g}}$-module as well. Therefore
$M_{2}=M_{1}\oplus
L_{{\bf g}}(\ell,\lambda)$. This is a contradiction. $\;\;\;\;\Box$

It has been proved in [DMZ] that if both $V^{1}$ and $V^{2}$ are
rational vertex operator algebras, then so is the tensor vertex operator
algebra $V^{1}\otimes V^{2}$.
In order to make this paper self-contained, we prove the following
special case.

{\bf Lemma 5.2.6.} {\it Let $V$ be a rational vertex operator superalgebra
and let $n$ be any positive integer. Then the tensor vertex operator
superalgebra $V\otimes {\bf F}^{n}$ is rational.}

{\sl Proof.} Let $M$ be any ${1\over 2}{\bf Z}$-graded weak $V\otimes
{\bf F}^{n}$-module, which is truncated from below.
Set $W=\{u\in M| (1\otimes L(-1))u=0\}$, where $L(-1)$ in the second
slot is from ${\bf F}^{n}$. Define a linear map:
\begin{eqnarray}
\psi :W\otimes {\bf F}^{n}\rightarrow M;\;
u\otimes a\mapsto a_{-1}u\;\;\;\mbox{for any }u\in W, a\in {\bf F}^{n}.
\end{eqnarray}
We shall prove that $M\simeq
W\otimes {\bf F}^{n}$ as a $V\otimes {\bf F}^{n}$-module.
It is clear that $\psi$ commutes with each $Y(a,z)$ for $a\in V$ and
it follows from Proposition 4.1.1 that $\psi$ is a ${\bf F}^{n}$-homomorphism.
Thus  $\psi$ is a $(V\otimes {\bf F}^{n})$-module homomorphism.
Since $M$ is a ${1\over 2}{\bf Z}$-graded weak
${\bf F}^{n}$-module truncated from below, by Proposition 4.3.5 $M$ is
a direct sum of copies of ${\bf F}^{n}$, so that $\phi$ is surjective.
Since ${\bf F}^{n}$ is simple, it follows from Proposition 4.1.1 that
for any $u\in W$, $\phi$ restricted to $u\otimes {\bf F}^{n}$ is a
injective, so that $\phi$ is injective on $W\otimes {\bf F}^{n}$.
Therefore, any $V\otimes {\bf F}^{n}$ has a canonical decomposition.
Since $W$ is a completely reducible $V$-module,
for proving that $W\otimes {\bf F}^{n}$ is a completely reducible
$V\otimes {\bf F}^{n}$-module, it is enough to prove that $M\otimes
{\bf F}^{n}$ is an irreducible $V\otimes {\bf F}^{n}$-module for any
irreducible $V$-module $M$. Let $M_{1}$ be any nonzero submodule of
$M\otimes {\bf F}^{n}$. Then there is a vector $0\ne u\in M_{1}$ such
that $(1\otimes L(-1))u=0$. Since ${\bf F}^{n}$ is simple, it follows from
the proof of Lemma 4.3 [L1] (see also [T]) that $\ker_{{\bf F}^{n}}
L(-1)={\bf C}{\bf 1}$. Therefore,
$\ker_{M\otimes {\bf F}^{n}}(1\otimes L(-1))=M\otimes {\bf 1}$. Then it is
easy to see that $u$ generates $M\otimes {\bf F}^{n}$ by $V\otimes
{\bf F}^{n}$. Thus $M_{1}=M\otimes {\bf F}^{n}$. Therefore, $M\otimes
{\bf F}^{n}$ is irreducible.$\;\;\;\;\Box$

{\bf Proposition 5.2.7.} {\it Let ${\bf g}_{0}$ be a finite-dimensional
simple Lie algebra  and let ${\bf g}_{1}$ is a finite-dimensional
${\bf g}_{0}$-module equipped with a nondegenerate symmetric bilinear
form $B_{1}$ such that}
\begin{eqnarray}
B_{1}(au.v)=-B_{1}(u,av)\;\;\;\mbox{ {\it for any} }a\in {\bf
g}_{0}, u,v\in {\bf g}_{1}.
\end{eqnarray}
{\it Let $\ell$ be such that $\ell-\Omega$ is a positive integer. Then the
vertex operator superalgebra $L_{{\bf g}}(\ell,0)$ is rational.}

{\sl Proof.} It is a simple consequence of Proposition 5.2.5,
Corollary 4.3.9 and Lemma 5.2.6.$\;\;\;\;\Box$

\newpage

\section{Generalized intertwining operators}
In [L2], we gave an analogue of the space of linear
homomorphisms from one module to another for a Lie algebra by
introducing the notion of generalized intertwining operators from one
module to another for a vertex operator algebra. In this section, we
generalize this result for vertex operator superalgebras.

Throughout this section, $V$ will be a fixed vertex
operator superalgebra.

\subsection{Generalized intertwining operators}
In this subsection, we define ``generalized intertwining operators'' from
one module to another for a vertex operator superalgebra $V$ and we
prove that the space of generalized intertwining operators has a
natural generalized $V$-module structure.

{\bf Definition 6.1.1}. Let $V$ be a vertex operator
superalgebra and let $M^{i}$ $(i=1,2)$ be $V$-modules. A formal
series
 $\phi (z)\in ({\rm Hom}_{{\bf C}}(M^{1},M^{2}))\{z\}$
is called an {\it even homogeneous generalized intertwining operator} if
it satisfies the following conditions:

(GIO1)\hspace{0.5cm}There are finitely many complex numbers
$h_{1},\cdots,h_{n}$ such that
\begin{eqnarray*}
& &\phi (z)\in \sum_{k=1}^{n}z^{h_{k}}\left({\rm Hom}_{{\bf C}}(M^{1},M^{2})
\right)[[z,z^{-1}]],\\
& &\phi (z) u\in \sum_{k=1}^{n}z^{h_{k}}\left(M^{2}\right)((z))
\;\;\;\mbox{for any }u\in M^{1}.
\end{eqnarray*}

(GIO2)\hspace{0.5cm}$[L(-1),\phi (z)]={d\over dz}\phi (z)=\phi '(z)$.

(GIO3)\hspace{0.5cm}$[L(0),\phi (z)]=h \phi (z)+z{d\over dz}\phi (z)$
for some complex number $h$, called the {\it weight} of $\phi (z)$.

(GIO4)\hspace{0.5cm}For any $a\in V$, there is a positive integer
$n$ such that
\begin{eqnarray}
(z_{1}-z)^{n}Y_{M^{2}}(a,z_{1})\phi (z)=(z_{1}-z)^{n}\phi
(z)Y_{M^{1}}(a,z_{1}).
\end{eqnarray}

Denote by $\left(G(M^{1},M^{2})\right)_{(h)}^{0}$ the vector space of
all even homogeneous generalized intertwining operators of weight $h$
from $M^{1}$ to $M^{2}$.

Similarly, a formal series $\phi (z)\in {\rm Hom}_{{\bf C}}(M^{1},M^{2})\{z\}$
is
called an {\it odd homogeneous generalized intertwining operator} if $\phi (z)$
satisfies (GIO1)-(GIO3) and the following condition:

${\rm (GIO4)}^{\prime}$\hspace{0.5cm}For any $({\bf Z}/2{\bf Z})$-homogeneous
$a\in
V$, there is a positive integer $n$ such that
\begin{eqnarray}
(z_{1}-z)^{n}Y_{M^{2}}(a,z_{1})\phi (z)=(-1)^{|a|}(z_{1}-z)^{n}\phi (z)
Y_{M^{1}}(a,z_{1}).
\end{eqnarray}

Denote by ($G(M^{1},M^{2}))_{(h)}^{1}$ the vector space of all
odd homogeneous generalized intertwining operators of weight $h$ from
$M^{1}$ to $M^{2}$. Then we set
\begin{eqnarray}
& &\left(G(M^{1},M^{2})\right)^{0}=\oplus_{h\in {\bf C}}\left(
G(M^{1},M^{2})\right)_{(h)}^{0},\\
& &\left(G(M^{1},M^{2})\right)^{1}=\oplus_{h\in {\bf C}}\left(
G(M^{1},M^{2})\right)_{(h)}^{1},\\
& &G(M^{1},M^{2})=(G(M^{1},M^{2}))^{0}\oplus (
G(M^{1},M^{2}))^{1}.
\end{eqnarray}
Any element of $\left(G(M^{1},M^{2})\right)^{0}$ is called an
{\it even generalized intertwining operator} and any
element of $\left(G(M^{1},M^{2})\right)^{1}$ is called an
{\it odd generalized intertwining operator}. Furthermore, we call any
element of
$G(M^{1},M^{2})$ a {\it generalized intertwining operator}.

It is clear that $G(M^{1},M^{2})$ is a $({\bf Z}/2{\bf
Z})$-graded vector space. If $\phi (z)$ is an even (resp. odd)
generalized intertwining operator (of weight $h$), then $\phi '(z)$ is
an even (resp. odd) generalized intertwining operator (of weight $h
+1$). Therefore, we have:
\begin{eqnarray}
& &{d\over dz}\cdot \left( G(M^{1},M^{2})\right)_{(h)}^{0}\subseteq
\left(G(M^{1},M^{2})\right)_{(h+1)}^{0},\\
& &{d\over dz}\cdot \left(G(M^{1},M^{2})\right)_{(h)}^{1}\subseteq
\left(G(M^{1},M^{2})\right)_{(h+1)}^{1}.
\end{eqnarray}
A generalized  intertwining operator $\phi(z)$ of weight $h$ is said
to be {\it primary} if it satisfies the following condition:
\begin{eqnarray}
[L(m),\phi(z)]=\left(z^{m+1}{d\over dz}+h(m+1)z^{m}\right)\phi(z)\;\;\mbox{for
any }m\in {\bf Z}.
\end{eqnarray}
A homogeneous (with respect to both gradings) generalized intertwining
operator $\phi (z)$ is
called a {\it lowest weight generalized intertwining operator} if satisfies the
following condition:
\begin{eqnarray}
(z_{1}-z_{2})^{k}Y_{M^{2}}(a,z_{1})\phi (z_{2})
=\varepsilon_{a,\phi}(z_{1}-z_{2})^{k}\phi (z_{2})Y_{M^{1}}(a,z_{1})
\end{eqnarray}
for any homogeneous element $a\in V$ and for any $k\ge {\rm wt} a$.

{\bf Definition 6.1.2}. For any homogeneous $a\in V, \phi (z)\in
G(M^{1},M^{2})$, we define
\begin{eqnarray}
& &Y(a,z_{0})\circ \phi (z_{2})\nonumber\\
&=&{\rm Res}_{z_{1}}
\left(z_{0}^{-1}\delta\left(\frac{z_{1}-z_{2}}{z_{0}}\right)Y_{M^{2}}(a,z_{1})
\phi
(z_{2})-\varepsilon_{a,\phi}z_{0}^{-1}\delta\left(\frac{z_{2}-z_{1}}{-z_{0}}
\right)\phi (z_{2})Y_{M^{1}}(a,z_{1})\right).\nonumber\\
\mbox{}
\end{eqnarray}
Then we extend the definition bilinearly to any $a\in V, \phi
(z)\in G(M^{1},M^{2})$.

{\bf Remark 6.1.3.}. For any $a\in V, \phi (z)\in
G(M^{1},M^{2})$, we set
\begin{eqnarray}
Y(a,z_{0})\circ \phi (z_{2})=\sum_{n\in{\bf Z}}a_{n}\circ \phi
(z_{2})z_{0}^{-n-1}.
\end{eqnarray}
It follows from (GIO4) and ${\rm(GIO4)}^{\prime}$ that $a_{n}\circ \phi
(z_{2})=0$ for $n$
sufficiently large.
By Remark 2.3.5, we have the following Jacobi identity:
\begin{eqnarray}
& &z_{0}^{-1}\delta\left(\frac{z_{1}-z_{2}}{z_{0}}\right)Y_{M^{2}}(a,z_{1})
\phi (z_{2})-\varepsilon_{a,\phi}z_{0}^{-1}\delta\left(\frac{z_{2}-z_{1}}
{-z_{0}}\right)\phi (z_{2})Y_{M^{1}}(a,z_{1})\nonumber \\
&=&z_{2}^{-1}
\delta\left(\frac{z_{1}-z_{0}}{z_{2}}\right)Y(a,z_{0})\circ \phi (z_{2}).
\end{eqnarray}

{\bf Proposition 6.1.4}. {\it For any homogeneous $a\in V, \phi (z)\in
G(M^{1},M^{2})$,  $a_{n}\circ \phi (z)$ is a homogeneous generalized
intertwining operator of weight $({\rm wt}a+{\rm wt}\phi -n-1)$ for
any $n$ and $|a_{n}\circ \phi (z)|=|a||\phi (z)|$.}

{\sl Proof}. This easily follows from the proofs of Lemma 3.1.8 and
Proposition 3.2.7.$\;\;\;\;\Box$

{\bf Lemma 6.1.5.} {\it Under Definition 6.1.2, we have}
\begin{eqnarray}
& &Y({\bf 1},z_{0})\circ \phi (z_{2})=\phi (z_{2})\;\;\mbox{ {\it for
any} }\phi (z_{2})\in G(M^{1},M^{2}),\\
& &L(-1)\circ \phi (z)\;(=\omega _{0}\circ \phi (z))={d\over dz}\phi
(z)\;\;\;\; \mbox{{\it for} }\phi (z)\in G(M^{1},M^{2}),\\
& & L(0)\circ \phi (z) (=\omega _{1}\circ \phi (z))= h \phi
(z)\;\;\;\;\mbox{{\it for} }\phi (z)\in G(M^{1},M^{2})_{(h)}.
\end{eqnarray}

{\sl Proof}. This directly follows from the definitions, (GIVO2) and (GIVO3).
$\;\;\;\;\Box$

{\bf Lemma 6.1.6.} For any $a\in V, \phi (z)\in G(M^{1},M^{2})$,
we have:
\begin{eqnarray}
[L(-1), Y(a,z_{0})]\circ \phi (z_{2})=Y(L(-1)a,z_{0})\circ \phi (z_{2})
={\partial\over\partial z_{0}}Y(a,z_{0})\circ \phi (z_{2}).
\end{eqnarray}

{\sl Proof.} It is similar to the proof of Lemma 3.1.7.$\;\;\;\;\Box$

In [L2], we have proved that $G(M^{1},M^{2})$ is a generalized
module for $V$ being a vertex operator algebra by using
FHL's notion of ``transpose'' intertwining operator [FHL]. Here, we
give a direct proof for $V$ being a vertex operator superalgebra.

{\bf Theorem 6.1.7}. {\it Under Definition 6.1.2, $
G(M^{1},M^{2})$ becomes a generalized $V$-module.}

{\sl Proof}. Notice that only the Jacobi identity is left to be checked. For
any homogeneous $a,b \in V,\phi (z)\in G(M^{1},M^{2})$, let $n$
be an positive integer such that
\begin{eqnarray}
& &(z_{3}-z)^{n}Y_{M^{2}}(a,z_{3})\phi
(z)=\varepsilon_{a,\phi}(z_{3}-z)^{n}\phi
(z)Y_{M^{1}}(a,z_{3}),\\
& &(z_{4}-z)^{n}Y_{M^{2}}(b,z_{4})\phi
(z)=\varepsilon_{b,\phi}(z_{4}-z)^{n}\phi
(z)Y_{M^{1}}(b,z_{4}),\\
& &(z_{3}-z_{4})^{n}Y_{M^{k}}(a,z_{3})Y_{M^{k}}(b,z_{4})
=\varepsilon_{a,b}(z_{3}-z_{4})^{n}Y_{M^{k}}(b,z_{4})Y_{M^{k}}(a,z_{3})
\end{eqnarray}
for $k=1,2$. By definition, we have
\begin{eqnarray}
& &Y(a,z_{1})\circ (Y(b,z_{2})\circ \phi (z))\nonumber\\
&=&{\rm Res}_{z_{3}}z_{1}^{-1}\delta\left(\frac{z_{3}-z}{z_{1}}\right)
Y_{M^{2}}(a,z_{3})(Y(b,z_{2})\circ \phi (z))\nonumber\\
& &-{\rm Res}_{z_{3}}z_{1}^{-1}\delta\left(\frac{z-z_{3}}{-z_{1}}\right)
\varepsilon_{a,b}\varepsilon_{a,\phi}(Y(b,z_{2})\circ \phi (z))
Y_{M^{1}}(a,z_{3})\nonumber\\
&=&{\rm Res}_{z_{3}}{\rm Res}_{z_{4}}(A-B-C+D)
\end{eqnarray}
where
\begin{eqnarray}
& &A=z_{1}^{-1}\delta\left(\frac{z_{3}-z}{z_{1}}\right)z_{2}^{-1}\delta
\left(\frac{z_{4}-z}{z_{2}}\right)Y_{M^{2}}(a,z_{3})Y_{M^{2}}(b,z_{4})\phi
(z),\\
& &B=z_{1}^{-1}\delta\left(\frac{z_{3}-z}{z_{1}}\right)z_{2}^{-1}\delta
\left(\frac{z-z_{4}}{-z_{2}}\right)\varepsilon_{b,\phi}Y_{M^{2}}(a,z_{3})\phi
(z)Y_{M^{1}}(b,z_{4}),\\
& &C=z_{1}^{-1}\delta\left(\frac{z-z_{3}}{-z_{1}}\right)z_{2}^{-1}\delta
\left(\frac{z_{4}-z}{z_{2}}\right)\varepsilon_{a,b}\varepsilon_{a,\phi}
Y_{M^{2}}(b,z_{4})\phi (z)Y_{M^{1}}(a,z_{3}),\\
& &D=z_{1}^{-1}\delta\left(\frac{z-z_{3}}{-z_{1}}\right)z_{2}^{-1}\delta
\left(\frac{z-z_{4}}{-z_{2}}\right)\phi (z)
\varepsilon_{a,b}\varepsilon_{a,\phi}\varepsilon_{b,\phi}
Y_{M^{1}}(b,z_{4})Y_{M^{1}}(a,z_{3}).
\end{eqnarray}
Similarly, we have
\begin{eqnarray}
& &Y(b,z_{2})\circ (Y(a,z_{1})\circ \phi (z))\nonumber \\
&=&{\rm Res}_{z_{3}}{\rm Res}_{z_{4}}(A'-B'-C'+D')
\end{eqnarray}
where
\begin{eqnarray}
& &A'=z_{1}^{-1}\delta\left(\frac{z_{3}-z}{z_{1}}\right)z_{2}^{-1}\delta
\left(\frac{z_{4}-z}{z_{2}}\right)Y_{M^{2}}(b,z_{4})Y_{M^{2}}(a,z_{3})\phi
(z),\\
& &B'=z_{1}^{-1}\delta\left(\frac{z_{3}-z}{z_{1}}\right)z_{2}^{-1}\delta
\left(\frac{z-z_{4}}{-z_{2}}\right)\varepsilon_{a,\phi}Y_{M^{2}}(b,z_{4})\phi
(z)Y_{M^{1}}(a,z_{3}),\\
& &C'=z_{1}^{-1}\delta\left(\frac{z-z_{3}}{-z_{1}}\right)z_{2}^{-1}\delta
\left(\frac{z_{4}-z}{z_{2}}\right)\varepsilon_{a,b}\varepsilon_{b,\phi}
Y_{M^{2}}(a,z_{3})\phi(z)Y_{M^{1}}(b,z_{4}),\\
& &D'=z_{1}^{-1}\delta\left(\frac{z-z_{3}}{-z_{1}}\right)z_{2}^{-1}\delta
\left(\frac{z-z_{4}}{-z_{2}}\right)\phi
(z)\varepsilon_{a,b}\varepsilon_{a,\phi}\varepsilon_{b,\phi}
Y_{M^{1}}(a,z_{3})Y_{M^{1}}(b,z_{4}).
\end{eqnarray}
By the property of $\delta$-function, we have
\begin{eqnarray}
z_{0}^{-1}\delta\left(\frac{z_{1}-z_{2}}{z_{0}}\right)z_{0}^{n}z_{1}^{n}
z_{2}^{n}Q=z_{0}^{-1}\delta\left(\frac{z_{1}-z_{2}}{z_{0}}\right)
(z_{3}-z)^{n}(z_{4}-z)^{n}(z_{3}-z_{4})^{n}Q
\end{eqnarray}
for any $Q\in \{A,B,C,D\}$. Thus
\begin{eqnarray}
& &z_{0}^{-1}\delta\left(\frac{z_{1}-z_{2}}{z_{0}}\right)z_{0}^{n}z_{1}^{n}
z_{2}^{n}(A-B-C+D)\nonumber \\
&=&z_{0}^{-1}\delta\left(\frac{z_{1}-z_{2}}{z_{0}}\right)(z_{3}-z)^{n}
(z_{4}-z)^{n}(z_{3}-z_{4})^{n}(A-B-C+D)\nonumber \\
&=&{\rm Res}_{z_{3}}{\rm Res}_{z_{4}}z_{0}^{-1}\delta\left(\frac{z_{1}-z_{2}}
{z_{0}}\right)z^{-1}\delta
\left(\frac{z_{3}-z_{1}}{z}\right)z^{-1}\delta\left(\frac{z_{4}-z_{2}}{z}
\right)\cdot X
\end{eqnarray}
where
$$X=(z_{3}-z)^{n}(z_{4}-z)^{n}(z_{3}-z_{4})^{n}Y_{M^{2}}(a,z_{3})
Y_{M^{2}}(b,z_{4})\phi (z).$$
Similarly, we have:
\begin{eqnarray}
& &z_{0}^{-1}\delta\left(\frac{z_{2}-z_{1}}{-z_{0}}\right)z_{0}^{n}z_{1}^{n}
z_{2}^{n}(A'-B'-C'+D')\nonumber \\
&=&{\rm Res}_{z_{3}}{\rm Res}_{z_{4}}z_{0}^{-1}\delta\left(\frac{z_{2}-z_{1}}
{-z_{0}}\right)z^{-1}\delta\left(
\frac{z_{3}-z_{1}}{z}\right)z^{-1}\delta\left(\frac{z_{4}-z_{2}}{z}\right)\cdot
X.
\end{eqnarray}
Therefore
\begin{eqnarray}
& &z_{0}^{n}z_{1}^{n}z_{2}^{n}\left(z_{0}^{-1}\delta\left(\frac{z_{1}-z_{2}}
{z_{0}}\right)Y(a,z_{1})Y(b,z_{2})-\varepsilon_{a,b}z_{0}^{-1}\delta\left(
\frac{z_{2}-z_{1}}
{-z_{0}}\right)Y(b,z_{2})Y(a,z_{1})\right)\phi (z)\nonumber \\
&=&{\rm Res}_{z_{3}}{\rm Res}_{z_{4}}z_{2}^{-1}\delta\left(\frac{z_{1}-z_{0}}
{z_{2}}\right)z^{-1}\delta\left(
\frac{z_{3}-z_{1}}{z}\right)z^{-1}\delta\left(\frac{z_{4}-z_{2}}{z}\right)
\cdot X.
\end{eqnarray}

On the other hand, we have
\begin{eqnarray}
& &z_{0}^{n}z_{1}^{n}z_{2}^{n}z_{2}^{-1}\delta\left(\frac{z_{1}-z_{0}}{z_{2}}
\right)Y(Y(a,z_{0})b,z_{2})\circ \phi (z)\nonumber\\
&=&{\rm Res}_{z_{4}}z_{0}^{n}z_{1}^{n}z_{2}^{n}z_{2}^{-1}\delta\left(\frac
{z_{1}-z_{0}}{z_{2}}\right)z_{2}^{-1}\delta\left(\frac{z_{4}-z}{z_{2}}\right)
Y_{M^{2}}(Y(a,z_{0})b,z_{4})\phi (z)\nonumber\\
& &-\varepsilon_{a,\phi}\varepsilon_{b,\phi}{\rm Res}_{z_{4}}
z_{0}^{n}z_{1}^{n}z_{2}^{n}z_{2}^{-1}\delta\left(\frac
{z_{1}-z_{0}}{z_{2}}\right)z_{2}^{-1}\delta\left(\frac{z-z_{4}}{-z_{2}}\right)
\phi (z)Y_{M^{1}}(Y(a,z_{0})b,z_{4}).\nonumber\\
\mbox{}
\end{eqnarray}
Since
\begin{eqnarray}
& &z_{0}^{n}Y_{M^{i}}(Y(a,z_{0})b,z_{4})\nonumber\\
&=&{\rm Res}_{z_{3}}z_{0}^{n}z_{0}^{-1}\delta\left(\frac{z_{3}-z_{4}}
{z_{0}}\right)Y_{M^{i}}(a,z_{3})Y_{M^{i}}(b,z_{4})\nonumber\\
& &-\varepsilon_{a,b}{\rm Res}_{z_{3}}z_{0}^{n}z_{0}^{-1}\delta\left(\frac
{z_{4}-z_{3}}
{-z_{0}}\right)Y_{M^{i}}(b,z_{4})Y_{M^{i}}(a,z_{3})\nonumber\\
&=&{\rm Res}_{z_{3}}z_{0}^{-1}\delta\left(\frac{z_{3}-z_{4}}
{z_{0}}\right)\left((z_{3}-z_{4})^{n}Y_{M^{i}}(a,z_{3})Y_{M^{i}}(b,z_{4})
\right)\nonumber\\
& &-\varepsilon_{a,b}{\rm Res}_{z_{3}}z_{0}^{-1}\delta\left(\frac
{z_{4}-z_{3}}{-z_{0}}\right)\left((z_{3}-z_{4})^{n}
Y_{M^{i}}(b,z_{4})Y_{M^{i}}(a,z_{3})\right)\nonumber\\
&=&{\rm Res}_{z_{3}}z_{4}^{-1}\delta\left(\frac{z_{3}-z_{0}}{z_{4}}\right)
\left((z_{3}-z_{4})^{n}Y_{M^{i}}(a,z_{3})Y_{M^{i}}(b,z_{4})\right),
\end{eqnarray}
we have
\begin{eqnarray}
& &z_{0}^{n}z_{1}^{n}z_{2}^{n}z_{2}^{-1}\delta\left(\frac{z_{1}-z_{0}}{z_{2}}
\right)Y(Y(a,z_{0})b,z_{2})\circ \phi (z)\nonumber\\
&=&{\rm Res}_{z_{3}}{\rm Res}_{z_{4}}z_{2}^{-1}\delta\left(\frac{z_{1}-z_{0}}
{z_{2}}\right)z_{2}^{-1}\delta\left(\frac{z_{4}-z}{z_{2}}\right)
z_{4}^{-1}\delta\left(\frac{z_{3}-z_{0}}{z_{4}}\right)\cdot X\nonumber\\
& &-\varepsilon_{a,\phi}\varepsilon_{b,\phi}{\rm Res}_{z_{3}}{\rm Res}_{z_{4}}
z_{2}^{-1}\delta\left(\frac{z_{1}-z_{0}}
{z_{2}}\right)z_{2}^{-1}\delta\left(\frac{z-z_{4}}{-z_{2}}\right)
z_{4}^{-1}\delta\left(\frac{z_{3}-z_{0}}{z_{4}}\right)\cdot X\nonumber\\
&=&{\rm Res}_{z_{3}}{\rm Res}_{z_{4}}
z_{2}^{-1}\delta\left(\frac{z_{1}-z_{0}}{z_{2}}\right)z^{-1}\delta\left(
\frac{z_{4}-z_{2}}{z}\right)z_{4}^{-1}\delta\left(\frac{z_{3}-z_{0}}{z_{4}}
\right)\cdot X.
\end{eqnarray}
Since
\begin{eqnarray*}
& &z_{2}^{-1}\delta\left(\frac{z_{1}-z_{0}}{z_{2}}\right)z^{-1}\delta\left(
\frac{z_{4}-z_{2}}{z}\right)z_{4}^{-1}\delta\left(\frac{z_{3}-z_{0}}{z_{4}}
\right)\\
&=&z_{2}^{-1}\delta\left(\frac{z_{1}-z_{0}}{z_{2}}\right)z_{4}^{-1}\delta\left(
\frac{z+z_{2}}{z_{4}}\right)z_{3}^{-1}\delta\left(\frac{z_{4}+z_{0}}{z_{3}}
\right)\\
&=&z_{1}^{-1}\delta\left(\frac{z_{2}+z_{0}}{z_{1}}\right)z_{4}^{-1}\delta\left(
\frac{z+z_{2}}{z_{4}}\right)z_{3}^{-1}\delta\left(\frac{z+z_{2}+z_{0}}{z_{3}}
\right)\\
&=&z_{1}^{-1}\delta\left(\frac{z_{2}+z_{0}}{z_{1}}\right)z_{4}^{-1}\delta\left(
\frac{z+z_{2}}{z_{4}}\right)z_{3}^{-1}\delta\left(\frac{z+z_{1}}{z_{3}}\right)
\\
&=&z_{2}^{-1}\delta\left(\frac{z_{1}-z_{0}}{z_{2}}\right)z^{-1}\delta\left(
\frac{z_{4}-z_{2}}{z}\right)z^{-1}\delta\left(\frac{z_{3}-z_{1}}{z}\right),
\end{eqnarray*}
then we obtain
\begin{eqnarray}
& &z_{0}^{n}z_{1}^{n}z_{2}^{n}\left(z_{0}^{-1}\delta\left(\frac{z_{1}-z_{2}}
{z_{0}}\right)Y(a,z_{1})Y(b,z_{2})-\varepsilon_{a,b}z_{0}^{-1}\delta\left(
\frac{z_{2}-z_{1}}
{-z_{0}}\right)Y(b,z_{2})Y(a,z_{1})\right)\phi (z)\nonumber \\
&=&z_{0}^{n}z_{1}^{n}z_{2}^{n}z_{2}^{-1}\delta\left(\frac{z_{1}-z_{0}}{z_{2}}
\right)Y(Y(a,z_{0})b,z_{2})\circ \phi (z).
\end{eqnarray}
Multiplying both sides by $z_{0}^{-n}z_{1}^{-n}z_{2}^{-n}$, we obtain
the Jacobi identity. $\;\;\;\;\Box$

{\bf Remark 6.1.7.} If $\phi(z)$ is a primary generalized intertwining
operator, then
$\phi(z)$ is a primary vector in the generalized $V$-module $
G(M^{1},M^{2})$, i.e., $L(m)\circ \phi (z)=0$ for any positive
integer $m$. If $\phi (z)$ is a lowest weight generalized
intertwining operator, then $a_{n}\circ \phi (z)=0$ for any
homogeneous $a\in V$ and for any $n\ge {\rm wt}a$.

\subsection{A universal property for $G(M^{1},M^{2})$}
In this subsection, we prove a universal property for $
G(M^{1},M^{2})$, by which we can identify the fusion rule of certain type
with the dimension of the space of $V$-homomorphisms from a certain
$V$-module to $G(M^{1},M^{2}))$.

Let $M$ be another $V$-module and let $\phi\in {\rm Hom}_{V}(M,
G(M^{1},M^{2}))$. Then we define a linear map $I_{\phi}(\cdot,z)$ by:
\begin{eqnarray}
\phi : & &M\rightarrow ({\rm Hom}_{{\bf
C}}(M^{1},M^{2}))\{z\}\nonumber\\
& &u\mapsto I_{\phi}(u,z)=\phi(u)(z)\;\;\;\mbox{ for }u\in M.
\end{eqnarray}
By definition, we have
\begin{eqnarray}
I_{\phi}(L(-1)u,z)=\phi(L(-1)u)(z)=L(-1)\circ \phi(u)(z)={d\over
dz}\phi(u)(z)={d\over dz}I_{\phi}(u,z).
\end{eqnarray}
Furthermore, for any homogeneous $a\in V, u\in M$, we have
\begin{eqnarray}
& &z_{0}^{-1}\delta\left(\frac{z_{1}-z_{2}}{z_{0}}\right)Y(a,z_{1})
I_{\phi}(u,z_{2})-\varepsilon_{a,u}z_{0}^{-1}\delta\left(\frac{z_{2}-z_{1}}
{-z_{0}}\right)I_{\phi}(u,z_{2})Y(a,z_{1})\nonumber\\
&=&z_{0}^{-1}\delta\left(\frac{z_{1}-z_{2}}{z_{0}}\right)Y(a,z_{1})
\phi(u)(z_{2})-\varepsilon_{a,u}z_{0}^{-1}\delta\left(\frac{z_{2}-z_{1}}
{-z_{0}}\right)\phi(u)(z_{2})Y(a,z_{1})\nonumber\\
&=&z_{2}^{-1}\delta\left(\frac{z_{1}-z_{0}}{z_{2}}\right)\left(Y(a,z_{0})
\circ \phi(u)(z_{2})\right)\nonumber\\
&=&z_{2}^{-1}\delta\left(\frac{z_{1}-z_{0}}{z_{2}}\right)\phi\left(
Y(a,z_{0})u\right)(z_{2})\nonumber\\
&=&z_{2}^{-1}\delta\left(\frac{z_{1}-z_{0}}{z_{2}}\right)I_{\phi}(
Y(a,z_{0})u,z_{2}).
\end{eqnarray}
This proves that $I_{\phi}(\cdot,z)$ is an intertwining operator of
type $\left(\begin{array}{c}M^{2}\\M,M^{1}\end{array}\right)$. If
$I_{\phi}(\cdot,z)=0$, then $\phi(u)(z)=0$ for all $u\in M$. Thus
$\phi=0$. Therefore, we obtain a linear injective map
\begin{eqnarray}
\pi : & &{\rm Hom}_{V}(M,{\rm G}(M^{1},M^{2}))\rightarrow I\left(\begin{array}
{c}M^{2}\\M,M^{1}\end{array}\right)\nonumber\\
& &\phi\mapsto I_{\phi}(\cdot,z).
\end{eqnarray}
On the other hand, for any intertwining operator $I(\cdot,z)$ of type
 $\left(\begin{array}{c}M^{2}\\M,M^{1}\end{array}\right)$, it is clear
that $I(u,z)\in G(M^{1},M^{2})$ for any $u\in M$. Then we
obtain a linear map $f_{I}$ from $M$ to $G(M^{1},M^{2})$
defined by $f_{I}(u)=I(u,z)$. Tracing back the argument above, we see
that $f_{I}$ is a $V$-homomorphism such that $I_{f_{I}}=I(\cdot,z)$.
Therefore we have proved the following universal property:

{\bf Theorem 6.2.1}. {\it Let $M^{1}$ and $M^{2}$ be modules for a
given vertex operator superalgebra $V$.
Then for any $V$-module $M$ and any intertwining operator $I(\cdot,z)$
of type $\left(\begin{array}{c}M^{2}\\M,M^{1}\end{array}\right)$,
there is a unique $V$-homomorphism $\psi$ from $M$ to $G(M^{1},M^{2})$
such that $I(u,z)=\psi(u)(z)$ for $u\in M$.}

The following corollary of Theorem 6.2.1 is parallel to Proposition 3.2.13.

{\bf Corollary 6.2.2.} {\it Let $M^{i}$ $(i=1,2,3)$ be three modules for a
vertex operator superalgebra $V$. Then giving  an intertwining
operator of type $\left(\begin{array}{c}M^{3}\\M^{1},M^{2}\end{array}\right)$
is equivalent to giving a $V$-homomorphism from $M^{1}$ to $
G(M^{2},M^{2})$.}

{\bf Remark 6.2.3.} Let $M^{1}$ and $M^{2}$ be any two modules for
a vertex operator superalgebra $V$. Then for any set $A$ of
homogeneous generalized
intertwining operators from $M^{1}$ to $M^{2}$, there exists a
generalized $V$-module $M$ and an intertwining operator $I(\cdot,z)$
of type $\left(\begin{array}{c}M^{2}\\M,M^{1}\end{array}\right)$ such
that each element of $A$ can be considered as an intertwining operator
of type $I(\cdot,z)$ evaluated at a vector of $M$. Also notice that the
commutator formula (2.2.6) implies (GVO4) or ${\rm (GVO4)}^{\prime}$.

{\bf Remark 6.2.4}. Just as in Lie algebra theory, $
G(M^{1},M^{2})$ is closely related to
a tensor product of certain two modules ([HL], [L2]). Here, we won't
go in that direction.

{\bf Proposition 6.2.5}. {\it Let $M$ be a $V$-module. Then $
G(V,M)\simeq M$.}

{\sl Proof}. For any $({\bf Z}/2{\bf Z})$-homogeneous $u\in M$, we define
\begin{eqnarray}
\phi _{u}(z)&:&V\rightarrow M[[z,z^{-1}]];\nonumber\\
& &\phi _{u}(z)a=\varepsilon _{a,u}e^{zL(-1)}Y(a,-z)u\;\;\;\;\mbox{for
any homogeneous }a\in V.
\end{eqnarray}
By definition, we get
\begin{eqnarray*}
[L(-1),\phi _{u}(z)]a
&=&L(-1)e^{zL(-1)}Y(a,-z)u-e^{zL(-1)}Y(L(-1)a,-z)u\\
&=&\left({d\over dz}e^{zL(-1)}\right)Y(a,-z)u+e^{zL(-1)}\left({d\over
dz}Y(a,-z)u \right)\\
&=&{d\over dz}\phi _{u}(z)a.
\end{eqnarray*}
For any $b\in V$, there is a positive integer $k$ such that
\begin{eqnarray}
(z_{0}+z_{2})^{k}Y(b,z_{0}+z_{2})Y(a,z_{2})u=(z_{0}+z_{2})^{k}Y(Y(b,z_{0})
a,z_{2})u\end{eqnarray}
for any $a\in V$ where $k$ is independent of $a$. By definition, we have
\begin{eqnarray}
(z_{0}+z_{2})^{k}Y(b,z_{0}+z_{2})e^{z_{2}L(-1)}\phi _{u}(-z_{2})=
\varepsilon_{b,u}(z_{0}+z_{2})^{k}e^{z_{2}L(-1)}\phi _{u}(-z_{2})Y(b,z_{0})a.
\end{eqnarray}
By the conjugation formula (2.2.4), we get
\begin{eqnarray}
(z_{0}+z_{2})^{k}Y(b,z_{0})\phi _{u}(-z_{2})=\varepsilon_{b,u}
(z_{0}+z_{2})^{k}\phi _{u}(-z_{2})Y(b,z_{0}).
\end{eqnarray}
Thus $\phi _{u}(z)\in G(V,M)$ for any $u\in M$. By Definition
6.1.2, one can easily
 prove that the linear map $\phi$ from $M$ to $G(V,M)$ is a
$V$-homomorphism.

On the other hand, let $\psi (z)$ be any even or odd generalized
intertwining operator from $V$ to $M$. Write $\psi(z)$ as
\begin{eqnarray}
\psi (z)=\sum_{i=1}^{k}\psi _{i}(z)\;\;\mbox{ where }z^{h_{i}}\phi
_{i}(z)\in ({\rm Hom}_{{\bf C}}(V,M))[[z,z^{-1}]],
\end{eqnarray}
where $h_{i} (i=1,\cdots,k)$ are different modulo ${\bf Z}$.
It is easy to see that every $\psi _{i}(z)$ is an even or odd
generalized intertwining operator
again. Now we may assume that $\psi (z)=\sum_{n\in h+{\bf Z}}\psi
_{n}z^{-n-1}$ for some complex number $h$. From condition (GIO2), we
have
\begin{eqnarray}
[L(-1),\psi_{n}]=-n\psi _{n-1}\;\;\mbox{for any }n\in h+{\bf Z}.\end{eqnarray}
If $\psi _{n}{\bf 1}=0$ for some nonzero $n$, then
\begin{eqnarray}
\psi_{n-1}{\bf 1}=-{1\over n}(L(-1)\psi _{n}-\psi_{n}L(-1)){\bf
1}=0.\end{eqnarray}
If $h$ is not an integer, it follows from (GIO1) that $\psi _{n}{\bf
1}=0$ for all $n\in h+{\bf Z}$. Furthermore, we have $\psi (z)=0$
because $\{a\in V|\psi (z)a=0\}$, the annihilating ideal of $\psi (z)$
in $V$, contains ${\bf 1}$.
For the rest of the proof, we assume $h=0$.
By condition (GIO1), it follows that $\psi _{n}{\bf 1}=0$ for any
nonnegative $n$. Therefore $\psi (z){\bf 1}$ involves only
nonnegative powers of $z$. Let $\psi _{-1}{\bf 1}=u\in M$. By using
(GIO2), one can easily get $\psi (z){\bf 1}=e^{zL(-1)}u$.
For any
$a\in V$, let $k$ be a positive integer such that
\begin{eqnarray}
(z-z_{1})^{k}\psi (z)Y(a,z_{1})
=\varepsilon_{a,\psi}(z-z_{1})^{k}Y(a,z_{1})\psi (z).\end{eqnarray}
Then
\begin{eqnarray}
z^{k}\psi (z)a
&=&\lim _{z_{1}\rightarrow 0}(z-z_{1})^{k}\psi (z)Y(a,z_{1}){\bf 1}\nonumber\\
&=&\lim _{z_{1}\rightarrow 0}\varepsilon_{a,\psi}(z-z_{1})^{k}Y(a,z_{1})
\psi (z){\bf 1}\nonumber\\
&=&\lim _{z_{1}\rightarrow 0}\varepsilon_{a,\psi}(z-z_{1})^{k}Y(a,z_{1})
e^{zL(-1)}u\nonumber\\
&=&\lim _{z_{1}\rightarrow
0}\varepsilon_{a,\psi}(z-z_{1})^{k}e^{zL(-1)}Y(a,z_{1}-z)u\nonumber\\
&=&\varepsilon_{a,\psi}z^{k}e^{zL(-1)}Y(a,-z)u.\end{eqnarray}
Thus $\psi (z)a=\varepsilon_{a,\psi}e^{zL(-1)}Y(a,-z)u$. That is,
$\psi(z)=\phi_{u}(z)$.
Therefore $G(V,M)$ is isomorphic to $M$ as a $V$-module.$\;\;\;\;\Box$

{\bf Remark 6.2.6}. If $M=V$, then $V=G(V,V)$. That is, any
generalized intertwining operator is a vertex operator.

\end{document}